\documentclass[aps,preprint,amsmath,amssymb,11pt]{revtex4}
\usepackage[T1]{fontenc}
\usepackage{mathrsfs}
\usepackage{epsfig}
\usepackage{slashed}
\usepackage{graphicx}
\usepackage{epstopdf}
\usepackage{graphics}
\usepackage{subfigure,psfrag}
\usepackage{amssymb}
\usepackage{amsmath}
\usepackage{amsthm}
\usepackage{multirow}
\usepackage{color}
\usepackage{makecell}

\begin{document}

\title{Searching for the doubly-charged Higgs bosons in the Georgi-Machacek model at the electron-proton colliders}
\author{Hao Sun\footnote{Corresponding author: haosun@mail.ustc.edu.cn \hspace{0.2cm} haosun@dlut.edu.cn}}
\author{Xuan Luo}
\author{Wei Wei}
\author{Tong Liu}
\affiliation{
Institute of Theoretical Physics, School of Physics,
Dalian University of Technology,
No.2 Linggong Road, Dalian, Liaoning, 116024, P.R.China}

\begin{abstract}
The Georgi-Machacek model is one of many beyond Standard Model scenarios with
an extended scalar sector which can group under the custodial $\rm SU(2)_C$ symmetry.
There are 5-plet, 3-plet and singlet Higgs bosons under the classification
of such symmetry in addition to the Standard Model Higgs boson.
In this paper, we study the prospects for detecting the doubly-charged Higgs boson
($\rm H_5^{\pm\pm}$) through the vector boson fusion production at the electron-proton (ep) colliders.
We concentrate on our analysis through $\mu$-lepton pair production via pair of same-sign W bosons decay.
The $\rm H_5^{\pm\pm}$ discovery potentials at the ep colliders are presented.
\end{abstract}

\maketitle

\section{Introduction}

The 125 GeV Standard Model (SM)-like particle was observed
at the Large Hadron Collider (LHC)\cite{SMHiggs_ATLAS}\cite{SMHiggs_CMS}.
Even so, it may still too early to conclude it is the whole story that responsible for
the electroweak symmetry breaking (EWSB) and even the mass of all the elementary particles.
In fact, from a theoretical point of view, there is no fundamental reason
for a minimal Higgs sector, as occurs in the SM.
It is therefore motivated to consider additional Higgs representations
that may also contribute to the symmetry breaking, and by doing this, one may even hold answers
to some longstanding questions in particle physics, such as the origin of neutrino mass,
the identity of the dark matter, and may establish a relationship with a yet undiscovered sector, etc.
In order to extend the Higgs sector in the SM, one can add isospin singlet or isospin doublet directly
to the SM Higgs doublet, or even higher isospin multiplet. No matter which way to use,
the following two requirements from the experimental data should be taken into account:
one is that the electroweak $\rho$ parameter should be very close to unity and
the other is that the tree level flavor changing neutral current (FCNC) processes should be strongly suppressed.

Such extended Higgs sectors are presented in many beyond Standard Model (BSM) scenarios,
for instance, the two Higgs doublet model\cite{ExtendedHiggs_THDM},
the minimal supersymmetric model\cite{ExtendedHiggs_SUSY},
the left-right symmetric model\cite{ExtendedHiggs_LR},
the little Higgs model\cite{ExtendedHiggs_LH}
and the Georgi-Machacek model\cite{Georgi:1985nv}\cite{Gunion:1990dt}.
The Georgi-Machacek (GM) model is one scenario proposed in the mid 80s extending the SM Higgs sector
with a complex $\rm SU(2)_L$ doublet field $\phi$ (Y = 1), a real triplet field $\xi$ (Y = 0)
and a complex $\rm SU(2)_L$ triplet field $\chi$ (Y = 2), where Y is the hypercharge.
Compare with the other Higgs extended models, the GM model has some desirable features.
In the GM model, the scalar potential maintain custodial $\rm SU (2)_C$ symmetry at the tree level\cite{Chaow}
which can keep the electroweak parameter close to unity, thus consistent with the experimental data.
Through appropriate Yukawa couplings with the leptons, the GM model can endow the neutrinos
with naturally light Majorana masses via the well known Type-II Seesaw Mechanism\cite{seesaw_TPYEII}.
After symmetry breaking, the physical fields can be organized by their transformation properties
under the custodial $\rm SU(2)_C$ symmetry into a fiveplet, a triplet, and two singlets.
One of the singlets is the SM-like Higgs, whose tree level couplings to fermions
and vector bosons may be enhanced in comparison to the SM case.
The fiveplet and the additional singlet couple to the electroweak gauge bosons but not SM quarks
at the tree level, whereas the triplet couples to the quarks but not the gauge bosons,
thus can be studied via different production mechanisms.
The custodial fiveplet contains exotic scalars including $\rm H^{\pm\pm}_5$, $\rm H^\pm_5$ and $\rm H^0_5$.
The single charged custodial fiveplet Higgs ($\rm H^\pm_5$) couple to the electroweak gauge bosons
and provide a good testing ground for the detection of the $\rm H^\pm W^\mp Z$ vertex\cite{Kanemura_HWZ}.
The neutral custodial fiveplet Higgs ($\rm H^0_5$) has the same mass as $\rm H^\pm_5$
and couples to WW and ZZ both and can be used to test the mass degeneracy of
charged and neutral scalar bosons in the GM model\cite{ILC_sinh}.
The presence of the doubly-charged Higgs particle ($\rm H_5^{\pm\pm}$) is phenomenal also quite interesting.
As in the GM model, a larger vacuum expectation value (VEV) is allowed due to the custodial symmetry.
The doubly-charged Higgs boson can decay into a pair of same-sign W bosons which include $\rm H^{\pm\pm}W^\mp W^\mp$ couplings.
The magnitude of this coupling is proportional to the larger VEV,
thus leading phenomenologically prominent and interesting signatures at colliders.

There have been extensive phenomenological studies of the exotic Higgs bosons in the GM model in the literature.
Most of them are based on analysis at the proton-proton (pp) colliders \cite{GMLHC_C.W.Chiang0}-\cite{HVVD_GM_Godunov}
and some are at the electron-position ($\rm e^+e^-$) colliders \cite{ILC_sinh}\cite{GMee_C.W.Chiang0}.
In this work, we investigate the discover potential of the heavy doubly-charged custodial fiveplet Higgs
in the GM model. A related study on the fiveplet Higgs at the pp collider can be found in Ref.\cite{LHC_H5} and
references cited above.
Here we present our study on doubly-charged scalar at the electron-proton (ep) colliders.
Ep colliders are hybrids between the $\rm e^+e^-$
and pp colliders, which consist of a hadron beam with an electron beam.
They provide a cleaner environment compared to the pp colliders
and higher center-of-mass (CMS) energies to the $\rm e^+e^-$ ones.
Currently, the proposed ep collider is the Large Hadron Electron Collider (LHeC) \cite{LHeC_1}\cite{LHeC_2}\cite{LHeC_3},
which is a combination of 60 GeV electron beam and 7 TeV proton beam of the LHC.
It can deliver up to 100 $\rm fb^{-1}$ integrated luminosity per year at a CMS energy of around 1 TeV
and 1 $\rm ab^{-1}$ over its lifetime.
This may later be extended to Future Circular electron-hadron Collider (FCC-eh)\cite{FCC-eh_1},
which features a 60 GeV (or higher) electron beam
with the 50 TeV proton beam from the Future Circular hadron Collider (FCC-hh).
This would result in a CMS energy up to 3.5 TeV with comparable luminosities to the LHeC\cite{FCC-eh_2}.
The proposed ep colliders are often used as a target for new physics searches.
There are already some works\cite{LHeC_BSM_Lindner, LHeC_BSM_Fischer, LHeC_BSM_Subhadeep, LHeC_BSM_ShouHua, LHeC_BSM_Kumar}
that have been done in the content of new physics searches based on them.
At the ep colliders, the doubly-charged Higgs boson can be produced through the vector boson fusion mechanism
($\rm e^-p\to\nu_e H_5^{--}j$).
We concentrate on our analysis through $\mu$-lepton pair production through pair of same-sign W bosons decay
($\rm e^-p\to\nu_{e}(H_5^{--}\to W^-W^-)j\to\mu_{e}\bar\nu_\mu\bar\nu_\mu \mu^-\mu^- j$).
Our studies are performed at the detector level, thanks to the contributions from the LHeC working group,
making the detector level studies available at both LHeC and FCC-eh.

Our paper is organized as follows:
In Section 2 the GM model is described in detail.
The constraints on the GM model are also presented.
In section 3 we study the collider phenomenology of the doubly-charged fiveplet resonance at the ep colliders.
Results include the signal and background analysis and the doubly-charged Higgs boson discovery potentials
are presented here. Finally we make our conclusions in the last section.

\section{SETUP THE MODEL FRAMEWORK}

\subsection{Description of the Georgi-Machacek model}

In the GM model, two $\rm SU(2)_L$ isospin triplet scalar fields,
$\chi$ with hypercharge $\rm Y=2$ and $\xi$ with $\rm Y=0$,
are introduced to the Higgs sector in addition to the original SM $\rm SU(2)_L$ doublet $\phi$ with $\rm Y=1$.
As we said, in constructing on extended Higgs sector,
the following two requirements should be taken into account:
$\rho$ is very close to unity and FCNC is suppressed.
It is the custodial symmetry that ensures $\rho =1$ at the tree level.
So the scalar content of the theory should be organized in terms of the $\rm SU(2)_L\otimes SU(2)_R$ custodial symmetry.
In order to make this symmetry explicit, we write the doublet in the form of a bidoublet $\Phi$ and combine the triplets to form a bitriplet $\Delta$:
\begin{align}
\Phi=\left(
\begin{array}{cc}
\phi^{0*} & \phi^+ \\
-\phi^-    & \phi^0
\end{array}
\right),\quad
\Delta=\left(
\begin{array}{ccc}
\chi^{0*} & \xi^+ & \chi^{++} \\
-\chi^-    & \xi^0 & \chi^{+} \\
\chi^{--} & -\xi^- & \chi^{0}
\end{array}
\right), \label{Higgs_matrices}
\end{align}
where we use the phase convention for the scalar field components:
\begin{eqnarray*}
\chi^{--}=(\chi^{++})^*,\,\, \chi^{-}=(\chi^{+})^*,\,\, \xi^-= (\xi^+)^*,\,\, \phi^-= (\phi^+)^*~.
\end{eqnarray*}
The neutral components in Eq.(\ref{Higgs_matrices}) can be parameterized into real and imaginary parts according to
\begin{equation}
\rm \phi^0 \to \frac{\nu_{\phi}}{\sqrt{2}} + \frac{\phi^{r} + i \phi^{i}}{\sqrt{2}},
\rm \qquad
\rm \chi^0 \to \nu_{\chi} + \frac{\chi^{r} + i \chi^{i}}{\sqrt{2}},
\rm \qquad
\rm \xi^0 \to \nu_{\xi} + \xi^r,
\end{equation}
where $\nu_\phi$, $\nu_\chi$ and $\nu_\xi$ are the VEVs of $\phi$, $\chi$ and $\xi$, respectively.
The scalar kinetic terms of the Lagrangian is written as
\begin{equation}
\mathcal{L}_{\rm kin} = |D^{(\phi)}_{\mu}\phi|^2 + \frac{1}{2}|D^{(\xi)}_{\mu}\xi|^2 + |D^{(\chi)}_{\mu}\chi|^2~,
\end{equation}
which gives the interaction terms between the scalars and the electroweak gauge bosons.
Write out explicitly, we have
\begin{eqnarray}
\mathcal{L}_{\rm kin} &\supset& \rm
(\nu_{\phi} + \phi^{r})^2\left(
\frac{g^2}{4}W_{\mu}^+W^{-,\mu} + \frac{g^2 + g^{\prime\,2}}{8}Z_{\mu}Z^{\mu}
\right)\nonumber + (\nu_\chi + \xi^0)^2
\left(
g^2 W_{\mu}^+W^{-,\mu}
\right) \\&+& \rm
(\sqrt{2}\, \nu_\chi + \chi^{r})^2\left(
\frac{g^2}{2}W_{\mu}^+W^{-,\mu} + \frac{g^2 + g^{\prime\,2}}{2}Z_{\mu}Z^{\mu} \
\right)\,.
\end{eqnarray}
Distinctly, the gauge boson masses are given by
\begin{equation}
\rm M_W^2 \equiv \frac{g^2}{4}(\nu_{\phi}^2 + 4 \nu_\chi^2 +4\nu_\xi ^2 )~,~~~~
\rm M_Z^2 \equiv \frac{g^2 + g^{\prime\,2}}{4}(\nu_{\phi}^2 + 8\nu_\chi^2)~.
\end{equation}
By imposing the custodial symmetry, i.e. $\rm \rho_{tree}=1$, we have
\begin{equation}
\rm \rho = \frac{M_W^2}{M_Z^2 cos^2 \theta_W} = \frac{\nu_{\phi}^2 + 4 \nu_\chi^2 +4\nu_\xi ^2 }{\nu_{\phi}^2 + 8\nu_\chi^2} = 1,
\end{equation}
therefore the vaccum alignment satisfy equation $\nu_\chi=\nu_\xi \equiv \nu_\Delta$, owing to the custodial symmetry.
In addition, the masses of weak gauge bosons are given by the same forms as those in SM
\begin{equation}
\rm M_W^2 = \frac{1}{4}g^2 \nu^2 \quad \Rightarrow \quad \nu_{\phi}^2 + 8 \nu_{\chi}^2 \equiv \nu^2 \approx (246\ GeV)^2.
\end{equation}
For convenience we can parameterize the VEVs as
\begin{equation}
\rm c_H \equiv  \cos\theta_H = \frac{\nu_{\phi}}{\nu}, \qquad
\rm s_H \equiv  \sin\theta_H = \frac{2\sqrt{2}\,\nu_\chi}{\nu}~.
\end{equation}
Following the above notation, the Lagrangian of the EWSB sector is succinctly given by
\begin{align}
\rm
{\cal L} \ =&\rm \
\frac{1}{2} {\rm tr}[ (D^{\mu}\Phi)^{\dagger} D_{\mu}\Phi ]
\ + \
\frac{1}{2} {\rm tr}[ (D^{\mu}\Delta)^{\dagger} D_{\mu}\Delta ]
\ - \ V(\Phi, \, \Delta) ~,
\end{align}
where $\rm D_{\mu}$ denotes the covariant derivative for $\Phi$ or $\Delta$.
The potential term is given by
\begin{align}
\begin{split}
\rm
V(\Phi, \, \Delta) \ =& \ \frac{1}{2} m_1^2 \, {\rm tr}[ \Phi^{\dagger} \Phi ] +
\frac{1}{2} m_2^2 \, {\rm tr}[ \Delta^{\dagger} \Delta ]
 +  \lambda_1 \left( {\rm tr}[ \Phi^{\dagger} \Phi ] \right)^2
 +  \lambda_2 \left( {\rm tr}[ \Delta^{\dagger} \Delta ] \right)^2
\\
&\rm
 +  \lambda_3 {\rm tr}\left[ \left( \Delta^{\dagger} \Delta \right)^2 \right]
 +  \lambda_4 {\rm tr}[ \Phi^{\dagger} \Phi ] {\rm tr}[ \Delta^{\dagger} \Delta ]
 +  \lambda_5 {\rm tr}\left[ \Phi^{\dagger} \frac{\sigma^a}{2} \Phi \frac{\sigma^b}{2} \right]
                  {\rm tr}[ \Delta^{\dagger} T^a \Delta T^b]
\\
&\rm
 + \mu_1 {\rm tr}\left[ \Phi^{\dagger} \frac{\sigma^a}{2} \Phi \frac{\sigma^b}{2} \right]
                               (P^{\dagger} \Delta P)_{ab}
 + \mu_2 {\rm tr}[ \Delta^{\dagger} T^a \Delta T^b]
                               (P^{\dagger} \Delta P)_{ab} ~,
\end{split}
\label{potential}
\end{align}
where $\sigma$'s and $\rm T's$ are the $2\times2$ (Pauli matrices) and $3\times3$ matrix representations of the SU(2) generators, and
\begin{align}
\rm
P  &= \frac{1}{\sqrt{2}} \left( \begin{array}{ccc}
-1 & i & 0 \\
0 & 0 & \sqrt{2} \\
1 & i & 0
\end{array}
\right). \nonumber
\end{align}

After the symmetry breaking from $\rm SU(2)_L\times SU(2)_R$  to $\rm SU(2)_C$,
the scalar fields in the GM model can be classified into different representations
under the custodial symmetry transformation.
The scalar fields from doublet $\Phi$ is decomposed through
${\bf 2}\otimes {\bf 2} \to {\bf 3}\oplus{\bf 1}$ into a 3-plet and a singlet.
Those from the triplet $\Delta$ is decomposed through ${\bf 3}\otimes {\bf 3} \to {\bf 5}\oplus{\bf 3}\oplus{\bf 1}$
into a 5-plet, a 3-plet and a singlet.
Among these $\rm SU(2)_C$ multiplets, the 5-plet states directly become physical Higgs bosons,
{\it i.e.}, $\rm H_5=(H_5^{\pm\pm}, H_5^\pm, H_5^0$).
For the two 3-plets, one of the linear combinations corresponds to physical Higgs field,
$\rm H_3=(H_3^\pm,H_3^0$), and the other becomes the Nambu-Goldstone (NG) bosons,
$\rm G^\pm$ and $\rm G^0$, which are absorbed into the
longitudinal components of the $\rm W^\pm$ and $\rm Z$ bosons, respectively.
Among the neutral fields, we have two $\rm SU(2)_C$ CP-even singlets,
$\rm H^1_\Phi = \phi^r$ and $\rm H^1_\Delta = \sqrt{\frac{1}{3}}\xi^r + \sqrt{\frac{2}{3}}\chi^r$,
that mix through a mixing angle $\alpha$ to render two physical Higgs bosons:
\begin{equation}
\rm h=cos\alpha H^1_\Phi - sin\alpha H^1_\Delta, \ \ \ H^0_1=sin\alpha H^1_\Phi + cos\alpha H^1_\Delta,
\end{equation}
with one of the two mass eigenstates being identified as the discovered 125 GeV SM-like Higgs boson.
Because of the custodial symmetry, the different charged Higgs boson states
belonging to the same $\rm SU(2)_C$ multiplet are almost degenerate in mass,
subject to small mass splitting ($\rm \sim {\cal{O}}(100)\ MeV$) due to electromagnetic corrections.
We will ignore such small mass differences and denote the Higgs masses by
$\rm M_{H_5}$, $\rm M_{H_3}$, $\rm M_{H_1}$ and $\rm M_{h}$
for the physical 5-plet, 3-plet, heavy singlet and the SM-like Higgs boson.
The five dimensionless scalar couplings $\lambda_1$ - $\lambda_5$ in the GM model
can be expressed in terms of the physical Higgs masses and the mixing angles $\alpha$ and $\rm \theta_H$ as
\begin{eqnarray}
&&\rm \lambda_1=\frac{1}{8\nu^2c^2_H} \left( M_h^2 c^2_\alpha + M^2_{H_1}s^2_\alpha \right), \nonumber \\
&&\rm \lambda_2=\frac{1}{6\nu^2s^2_H} \left[ 2M^2_{H_1}c^2_\alpha+2M^2_h s^2_\alpha+3M_2^2-2M^2_{H_5}+6c^2_H(M^2_{H_3}-M^2_1) \right], \nonumber\\
&&\rm \lambda_3=\frac{1}{\nu^2 s^2_H} \left[ c^2_H \left( 2M^2_1-3M^2_{H_3} \right) + M^2_{H_5}-M^2_2 \right], \nonumber \\
&&\rm \lambda_4=\frac{1}{6\nu^2s_H c_H} \left[ \sqrt{6} s_{\alpha} c_\alpha \left( M^2_h-M^2_{H^0} \right) + 3s_H c_H \left( 2M^2_{H_3}-M^2_1 \right) \right], \nonumber \\
&&\rm \lambda_5=\frac{2}{\nu^2} \left( M^2_1-M^2_{H_3} \right),
\end{eqnarray}
where $\rm c_\theta$ and $\rm s_\theta$ are abbreviations for $\rm cos\theta$ and $\rm sin\theta$
for $\rm \theta=\alpha, \theta_H$, respectively, and $\rm M_1$ and $\rm M_2$ are defined as
\begin{eqnarray}
\rm M_1^2 = -\frac{\nu}{\sqrt{2} s_H}\mu_1,\ \ \ M_2^2 =-3\sqrt{2} s_H \nu \mu_2.
\end{eqnarray}
The Feynman rules for the vertices involving a scalar and two gauge bosons are defined as $\rm ig_s V V g^{\mu\nu}$. The couplings are given by
\begin{eqnarray}
&\rm g_{hW^+W^{+*}}  = c_W^2 g_{hZZ} = -\frac{e^2}{6 s_W^2} \left( 8\sqrt{3} s_\alpha \nu_\chi - 3 c_\alpha \nu_\phi\right),  \nonumber \\
&\rm g_{HW^+W^{+*}}  = c_W^2 g_{HZZ} =   \frac{e^2}{6 s_W^2} \left( 8\sqrt{3} c_\alpha \nu_\chi + 3 s_\alpha \nu_\phi\right), \nonumber \\
&\rm g_{H_5^0 W^+ W^{+*}} = \sqrt{\frac{2}{3}} \frac{e^2}{s_W^2} \nu_{\chi},  \nonumber \\
&\rm g_{H_5^0 Z Z} = -\sqrt{\frac{8}{3}} \frac{e^2}{s_W^2 c_W^2} \nu_{\chi}, \nonumber \\
&\rm g_{H_5^+ W^{+*} Z} =   -\frac{\sqrt{2} e^2 \nu_{\chi}}{c_W s_W^2}   , \nonumber \\
&\rm g_{H_5^{++} W^{+*} W^{+*}} =   \frac{2 e^2 \nu_{\chi}}{s_W^2}.
\end{eqnarray}
The full feynman rules can be found, for example, in ref.\cite{GMdecoupling},
where in our analysis, we concentrate on the study of $\rm H_5^{\pm\pm} W^\mp W^\mp$ vertex.

\subsection{Constraints on the parameter space of the GM model}

\subsubsection{Theoretical constraints}

The theoretical constraints on the parameters of GM model,
such as the unitarity of the perturbation theory and stability of the electroweak vacuum,
had been considered in \cite{GMdecoupling}.

The $2\to 2$ scalar scattering matrixs can be expanded in terms of the Legendre polynomials:
\begin{equation}
\rm \mathcal{A} = 16\pi\sum_J (2J+1) a_J P_J(\cos\theta),
\end{equation}
where $\rm J$ is the (orbital) angular momentum and $\rm P_J(\cos\theta)$ are the Legendre polynomials.
Perturbative unitarity requires that the zeroth partial wave amplitude, $\rm a_0$,
satisfy $\rm |a_0| \leq 1$ or $\rm |{\rm Re} \, a_0| \leq \frac{1}{2}$.
In the high energy limit, only those diagrams involving the four-point scalar couplings
contribute to $2 \to 2$ scalar scattering process,
all diagrams involving scalar propagators are suppressed by the square of the collision energy.
Therefore the dimensionful couplings $\rm M_1$, $\rm M_2$, $\mu_2^2$, and $\mu_3^2$
are not constrained directly by perturbative unitarity.
Perturbative unitarity provides a set of constraints on the parameters of the scalar potential \cite{GMdecoupling},
\begin{eqnarray}
\rm \sqrt{P_{\lambda}^2 + 36\lambda_2^2} + |6\lambda_1 +7\lambda_3 + 11\lambda_4| &<& 4\pi~,\\
\rm \sqrt{Q_{\lambda}^2 +\lambda_5^2} + |2\lambda_1 -\lambda_3 +2\lambda_4 | &<& 4\pi~,\\
\rm |2\lambda_3 + \lambda_4|   &<& \pi~,\\
\rm |\lambda_2 - \lambda_5|   &<& 2\pi~,
\end{eqnarray}
with $\rm P_{\lambda} \equiv 6\lambda_1 -7\lambda_3 -11\lambda_4$,
$\rm Q_{\lambda} \equiv 2\lambda_1 +\lambda_3 -2\lambda_4$.
In addition
\begin{equation}\label{eq:25}
\lambda_2 \in \left(-\frac{2}{3}\pi, \frac{2}{3}\pi\right)~,~~~~
\lambda_5 \in \left(-\frac{8}{3}\pi, \frac{8}{3}\pi\right)~.
\end{equation}

Another constraint comes from the stability of electroweak vacuum,
i.e., the potential must be bounded from below. This requirement restricts $\lambda_{3,4}$ to satisfy
\begin{equation}\label{eq:34}
\lambda_3 \in \left(-\frac{1}{2}\pi, \frac{3}{5}\pi\right)~,~~~~
\lambda_4 \in \left(-\frac{1}{5}\pi, \frac{1}{2}\pi\right)~.
\end{equation}

\subsubsection{Experimental constraints}

Various measurements of the SM quantities in experiments provide stringent constraints on the parameters of GM model.
These include, for example,

\begin{itemize}
 \item Modification of the SM-like higgs couplings \cite{ATLAS_CMSHiggs}.
The Higgs coupling in the GM model depends on the triplet VEV $\rm \nu_\chi$
and the mixing angle of the two singlets $\alpha$.
The measured signal strengths of the SM-like higgs in various channels
from a combined ATLAS and CMS analysis of the LHC pp collision data
at $\rm \sqrt{s}$ = 7 and 8 TeV constrain $\rm \nu_\chi$ and $\alpha$ in the GM model severely.
 \item Electroweak precision tests \cite{indirectc}.
The presence of additional scalar states, charged under the EW symmetry,
generates a non-zero contribution to the oblique parameters $\rm S$ and $\rm T$~\cite{Peskin:1991sw}.
Setting $\rm U = 0$, the experimental values for the oblique parameters $\rm S$ and $\rm T$
are extracted for a reference SM Higgs mass $\rm M_h^{\rm SM} = 125\ GeV$ as $\rm S_{exp} = 0.06\pm 0.09$
and $\rm T_{\rm exp} =0.10\pm 0.07$ with a correlation coefficient of $\rm \rho_{ST} = +0.91$.
\end{itemize}

\section{PROCESS ANALYSIS AND NUMERICAL CALCULATIONS}

\subsection{Signal and Background Analysis}

\begin{figure}[hbtp]
\includegraphics[scale=0.5]{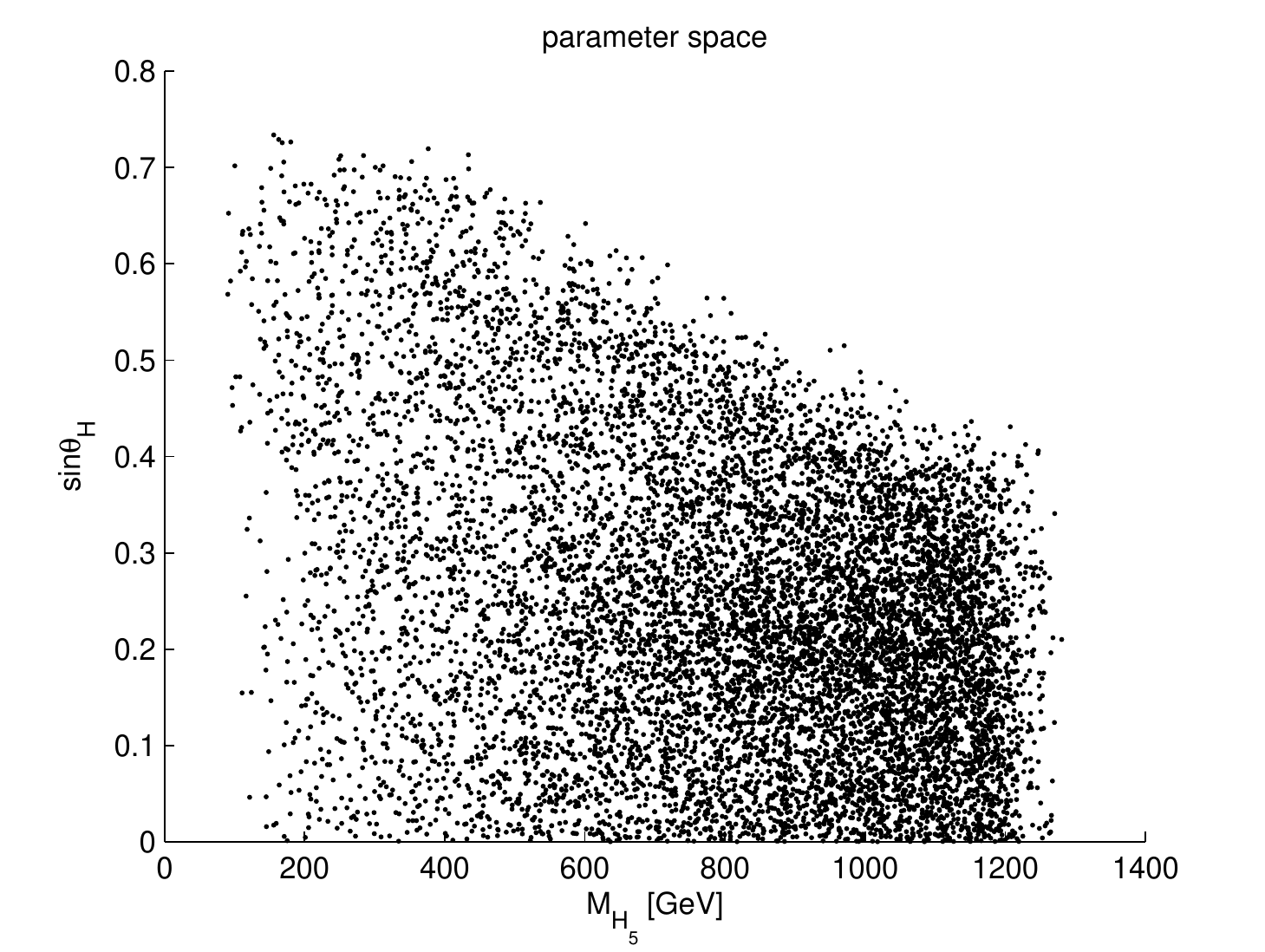}
\includegraphics[scale=0.26]{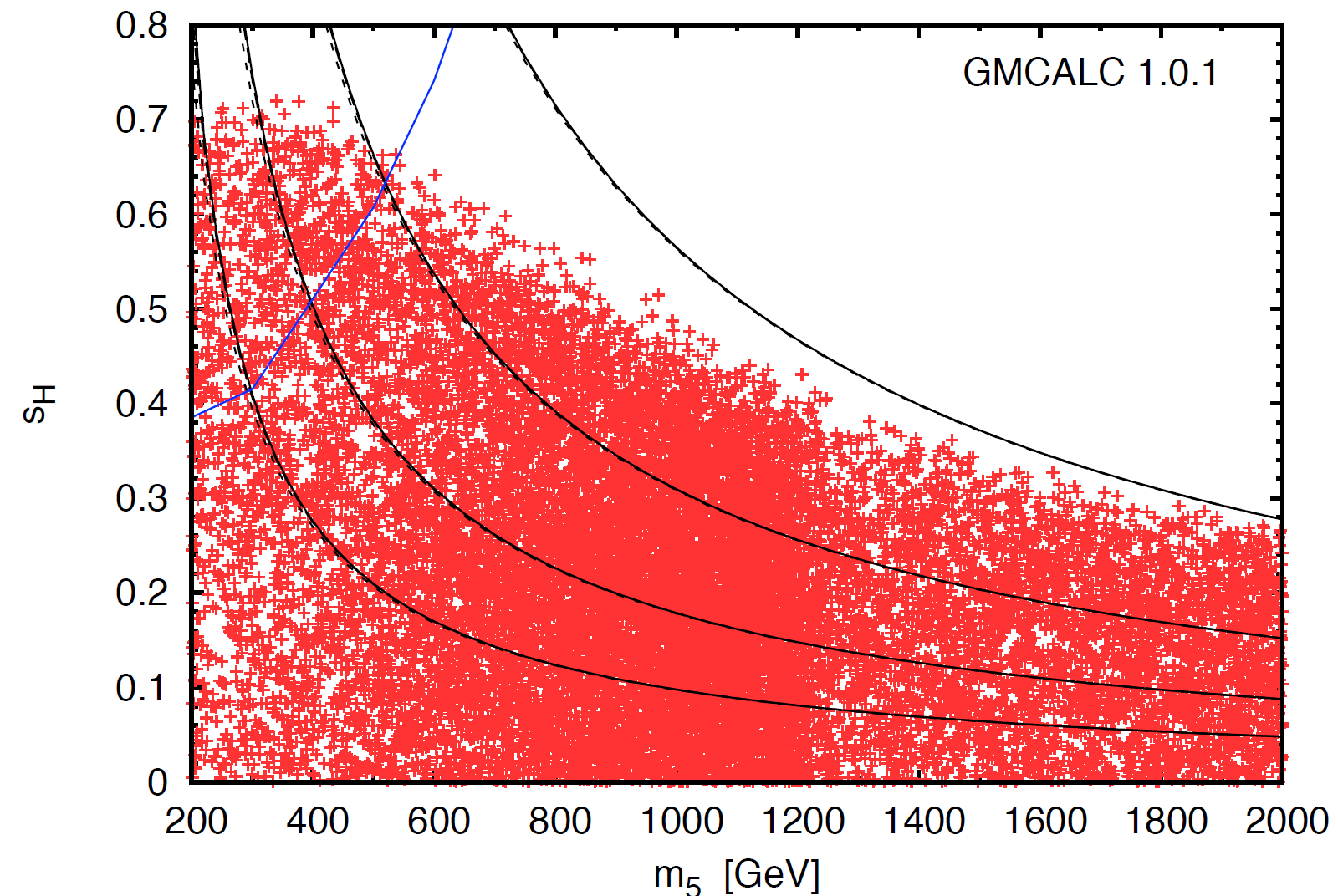}
\caption{\label{allowed parameter space}
Left: the scatter point plot of the allowed parameter space in the plane of $\rm sin\theta_H$ and $\rm M_{H_5}$.
Right: the same one, but taken from Ref.\cite{LHC_H5} including the LHC experimental boundary.}
\end{figure}

\begin{figure}[hbtp]
\includegraphics[scale=0.5]{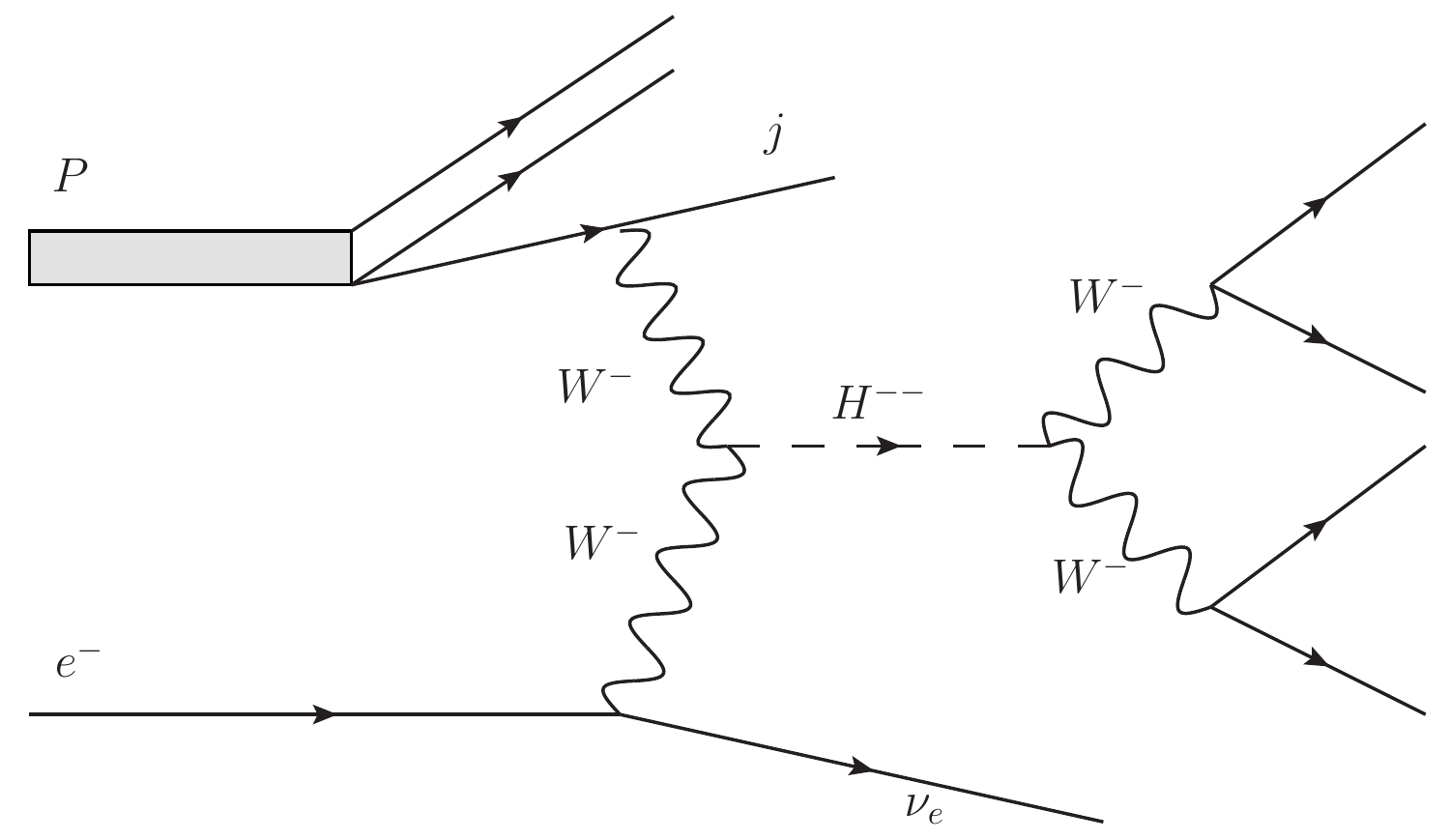}
\caption{\label{Signal Feynman diagram}
Feynman diagram for the signal process $\rm e^- p \to \nu_e (H_5 ^{--}\to W^- W^- ) j $ with
$\rm \nu_e \bar\nu_\ell \bar\nu_\ell \ell^- \ell^- j$, $\rm \nu_{e} \bar\nu_\ell \ell^- jjj$ and $\rm jjjjj$
decay modes at the ep colliders.}
\end{figure}

\begin{figure}[hbtp]
\includegraphics[scale=0.4]{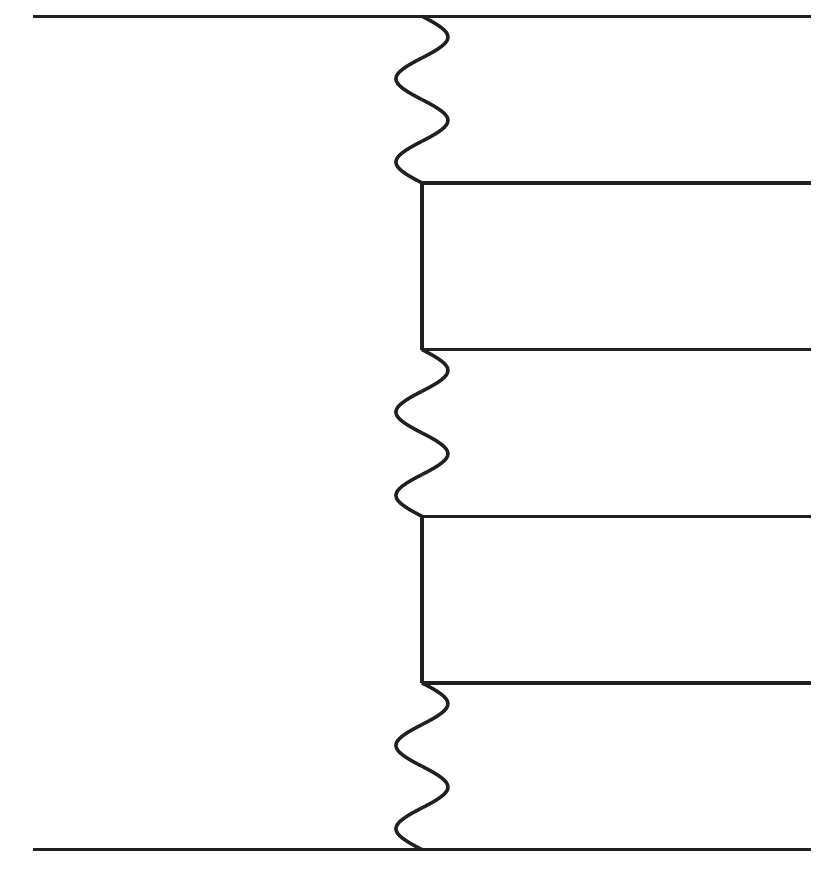}
\includegraphics[scale=0.4]{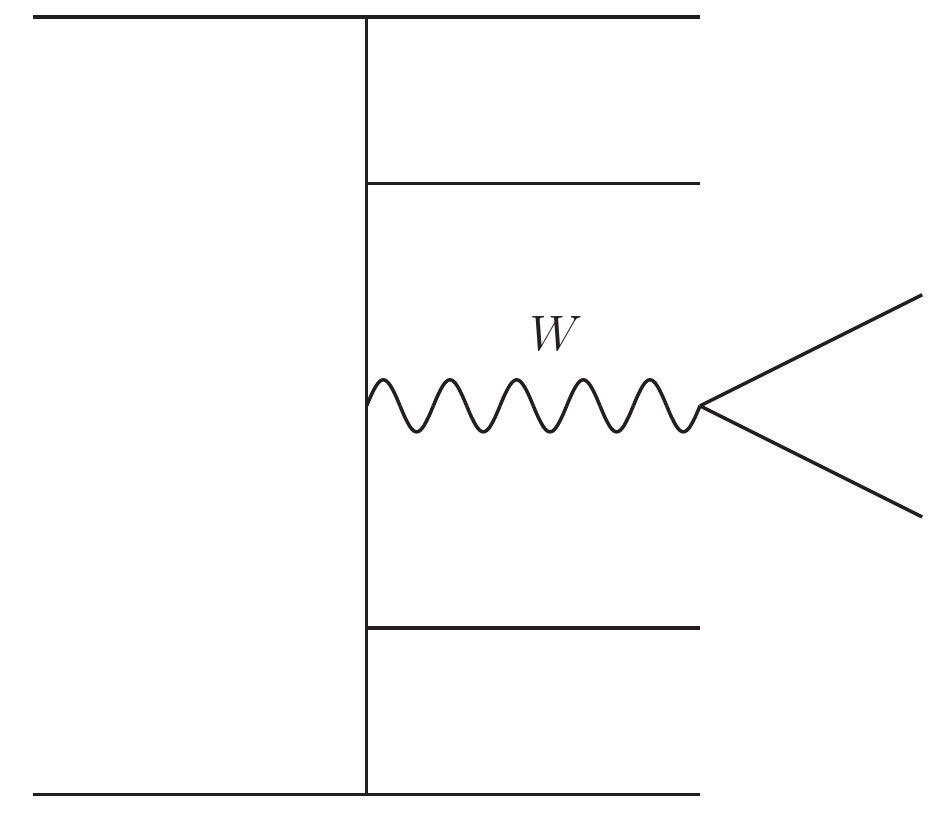}
\includegraphics[scale=0.4]{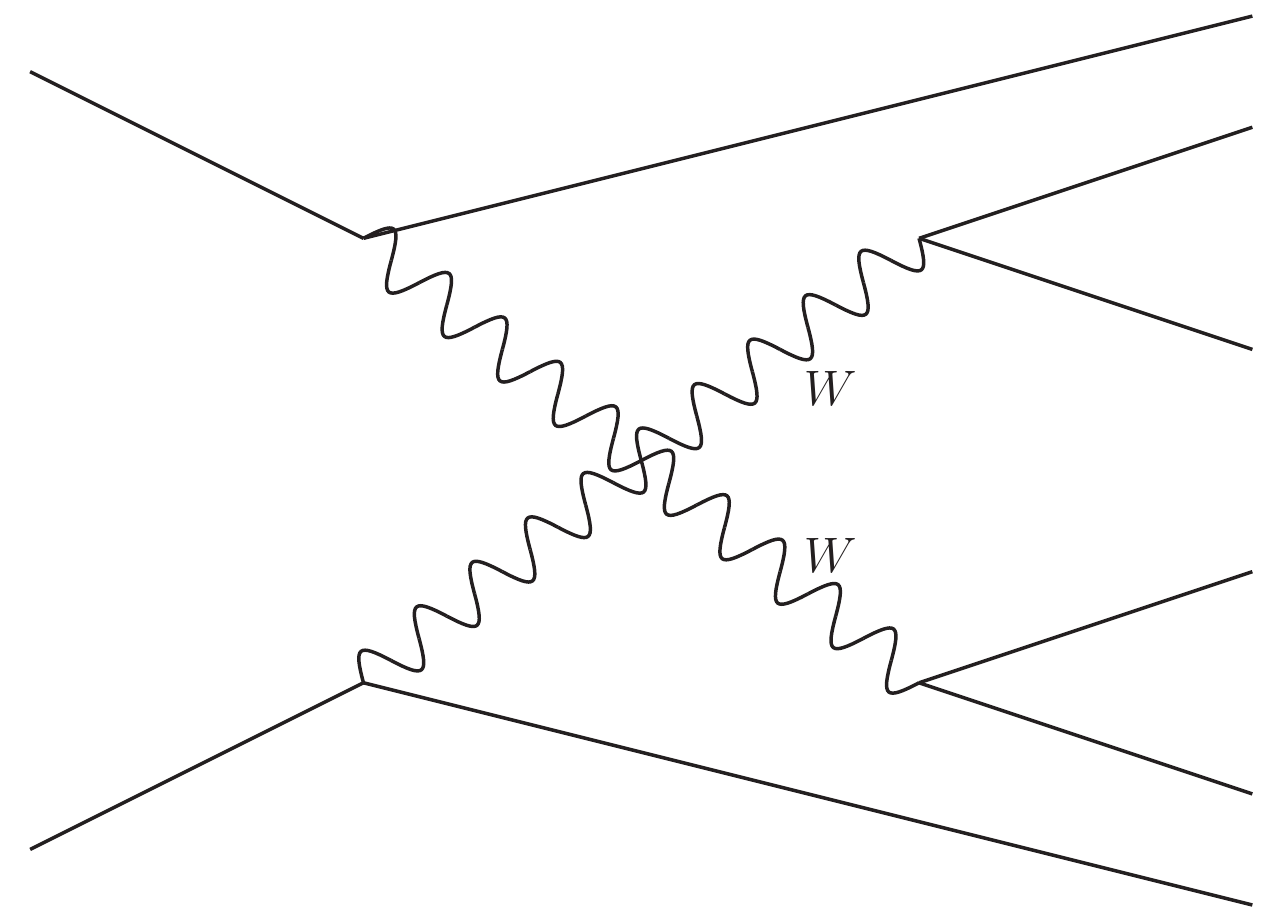}
\caption{\label{feyn_background}
Illustrated Feynman diagrams for the 0-$\rm W^-$ (non-resonant), 1-$\rm W^-$ and 2-$\rm W^-$ background processes respectively.}
\end{figure}

Our purpose is to study the prospects for detecting the doubly-charged
Higgs boson ($\rm H_5^{\pm\pm}$) in the GM model at the ep colliders.
Through ep collision, the fiveplet scalar states are produced
through the vector boson fusion (VBF), Drell-Yan or associated production ($\rm VH_5$) modes.
For the doubly-charged sector, the promising channel is via the same-sign VBF mechanism
\begin{eqnarray}
 \rm e^-p \to \nu_{e} (W^-W^- \to H_5^{--}) j,
\end{eqnarray}
where j refers to jets with $\rm j=u,c,\bar{d},\bar{s}$.
Considering the decays of the doubly-charged Higgs bosons at the tree level,
there are mainly three types\cite{GMee_C.W.Chiang0}. One is the scalar-gauge-gauge interaction which includes
$\rm H^{\pm\pm}_5\to W^\pm W^\pm$ decay mode, the second type is scalar-scalar-gauge interaction
which includes $\rm H^{\pm\pm}_5\to W^\pm H^\pm_3$ decay mode
and the third one is scalar-scalar-scalar interaction which includes $\rm H^{\pm\pm}_5\to H^\pm_3 H^\pm_3$ decay mode.
Notice when the mass of $\rm H_5^{\pm\pm}$ is smaller than the total mass of the final states,
one or both of gauge bosons must be off-shell.
So that the above decays should be understood to have 3- or 4-body final states.
In Fig.\ref{allowed parameter space} we present the scatter plot of the allowed parameter space of GM model
in the plane of $\rm sin\theta_H$ and $\rm M_{H_5}$.
This scan was done by the public available package GMCALC\cite{Hartling:2014xma}
by applying the theoretical and experimental constraints include
the perturbation unitarity of the scalar couplings, the stability of the electroweak vacuum,
the S parameter, as well as the indirect constraints from $\rm b\to s\gamma$.
The scan can be checked with the similar one in Ref.\cite{LHC_H5},
presented in the right panel in Fig.\ref{allowed parameter space}.
The blue curve in this plot indicates the LHC experiment limit from Ref\cite{GM_Boundary_a},
where the authors use a recasting of an ATLAS like-sign wwjj cross-section measurement at the 8 TeV LHC
in the context of the GM model, and find an exclusion of the doubly-charged member
for $\rm H_5$ values of about 140 to 400 GeV at $\rm \sin\theta_H$ = 0.5\cite{GM_Boundary_a}.
The points above the curve are excluded.
From the plots we can see that $\rm M_{H_5}$ spans a wide range
while $\rm sin\theta_H$ is constrained correspondingly.
More specifically, we see that when $\rm M_{H_5}$ is less than 600 GeV,
middle and large values of $\rm \sin\theta_H$ are ruled out by the data.
For the area where $\rm M_{H_5}$ is larger than 600 GeV,
$\rm \sin\theta_H$ is mainly constrained from the other constraints.
For most of the scan points, $\rm BR(H^{\pm\pm}_5 \to W^\pm W^\pm)$ were close to unity.
Therefore, for simplicity, the $\rm H^{\pm\pm}_5$ state can be assumed to decay entirely into vector boson
pairs for masses above the same-sign W pair threshold\cite{LHC_H5}.
This assumption holds in the vast majority of the GM model parameter space.
We thus concentrate on the production of doubly-charged Higgs boson via same-sign W bosons decay mechanism.
We should notice that the CMS collaboration has recently reported the most stringent limits
on the production of doubly charged Higgs bosons through electroweak production of same-sign W
boson pairs in the two jet and two same-sign lepton final
state in proton-proton collisions at 13 TeV\cite{GM_Boundary_b}.
We also include this limit in our numerical results where more details will shown in the following.
Considering W boson decays, we can in principle study three decay modes:
one is the same-sign leptonical decay mode (mode A), the second is the
semileptonical decay mode (mode B) and the third one is the multi-jet decay mode (mode C):
\begin{eqnarray}\nonumber
&&\rm mode\ A:\ \ e^-p \to \nu_{e} ( H_5^{--} \to W^-W^-) j \to \nu_{e} \bar\nu_\ell \bar\nu_\ell \ell^- \ell^- j \\\nonumber
&&\rm mode\ B:\ \ e^-p \to \nu_{e} ( H_5^{--} \to W^-W^-) j \to \nu_{e} \bar\nu_\ell \ell^- jjj \\
&&\rm mode\ C:\ \ e^-p \to \nu_{e} ( H_5^{--} \to W^-W^-) j \to \nu_{e} jjjjj.
\end{eqnarray}
Notice here j refers to jets with $\rm j=u,c,\bar{d},\bar{s}$,
and other possibilities from W boson decay, $\rm \ell=e,\mu$ are the leptons.
The corresponding feynman diagrams are plotted in Fig.\ref{Signal Feynman diagram}.
The large multi-jet backgrounds make modes B and C the challenge choices to detect the doubly-charged Higgs bosons.
We thus focus on the same-sign VBF of $\rm H_5 ^{--}$ production with same-sign leptonical decay modes.
The reasons we focus on it are listed as follows:
\begin{itemize}
 \item The SM background for events with two same-sign leptons
 is much suppressed compared to those with opposite-sign leptons or only one lepton, thus provide us a clean channel.
 \item This $\rm H_5 ^{--}$ production mechanism is dominant
 if its mass is above $\sim$ 300 GeV and $\nu_\triangle \gtrsim$ 10 GeV, thus have some kinematical advantages.
 \item In this $\rm H_5 ^{--}$ production process,
 leptons (more specifically, $\mu$-lepton, and the reason will present in the following)
 possibly arise only from the decay of the singly-produced $\rm H_5 ^{--}$.
 This fact allows a less parameter-dependent estimation of the acceptance
 and efficiency of events with two same-sign leptons, in comparison with the Drell-Yan production
 of $\rm H_5^{\pm\pm} H_5^{\mp\mp}$ or $\rm H_5 ^{\pm\pm} H_3 ^{\mp}$ and the associated production of $\rm H_5 ^{\pm\pm} W^{\mp}$.
\end{itemize}
Within mode A, by distinguishing $\rm e,\ \mu$ leptons,
we can have $\rm e^-e^-$, $\rm e^-\mu^-$ or $\rm \mu^-\mu^-$ final combinations of the observables.
At a first glance, in a very roughly estimation,
we can calculate one and times three to get them all, in order to get more statistic samples.
However this is not true in our case, since what we are considering is the ep colliders,
whose initial collision particle is electron, and will be much easier to appear in the final state.
For $\rm e^-e^-$ and $\rm e^-\mu^-$ combinations, they will be affected
by the $\rm e^- p \to e^- W^- j$ background, whose cross section is about 1.106 pb, which is unfortunately quite large.
Even after suitable veto cuts this background can still be two orders larger than the signal.
In contrast, $\rm \mu^-\mu^-$ final state will have much clean environment and become the advantaged choice
to detect the doubly-charged Higgs boson at the ep colliders, say,
\begin{eqnarray}
\rm e^-p \to \nu_{e} ( H_5^{--} \to W^-W^-) j \to \nu_{e} \bar\nu_\mu \bar\nu_\mu \mu^- \mu^- j.
\end{eqnarray}
In this case, the studied topology is characterized by a same-sign $\mu$-lepton pair,
a forward rapidity jet, plus missing transverse momentum from the undetected neutrinos.
The only background is simply $\rm e^-p\to\nu_e \mu^-\mu^-\nu_\mu\nu_\mu j$,
including 0-$\rm W^-$ (non-resonant), 1-$\rm W^-$ and 2-$\rm W^-$ (resonant) contributions.
Their illustrated diagrams are given in Fig.\ref{feyn_background} for clarity.
By our numerical confirmation,
we find that, it is exactly the 2-$\rm W^-$ resonant contributions ($\rm e^-p\to\nu_e (W^-W^-\to \mu^-\nu_\mu\mu^-\nu_\mu)j$)
that actually dominant over the other ones and stands more than $90\%$ of the total contributions.
This is also the direct reason why the background distributions
are distinguishable from those of the signal, see our discussion bellow.

\subsection{Simulation}

For the simulation of the collider phenomenology,
we use FeynRules\cite{FeynRules2.0} to extract the Feynman Rules from the Lagrangian.
The model is generated into Universal FeynRules Output(UFO) files\cite{UFO}
and then fed to the Monte Carlo event generator MadGraph@NLO\cite{MadGraph5} for the generation of event samples.
We pass the generated parton level events on to PYTHIA6.4\cite{Pythia}
which handles the initial and final state parton shower, hadronization, heavy hadron decays, etc.
Then, we pass the events on to Delphes3.4.1\cite{Delphes} 
which handles the detector effect\footnote{For the detector simulations, 
we used a dedicated Delphes card file for the LHeC and FCC-eh detector designs from the
LHeC/FCC-eh Higgs physics study group. This card file has been tested and verified by the Higgs group 
and can be obtained by request.}.
We use NN23LO1\cite{Ball:2012cx}\cite{Deans:2013mha} parton distribution functions for all event generations.
The factorisation and renormalisation scales for both the signal and the background simulation
are done with the default MadGraph5 dynamic scales.
The electron polarization is assumed to be unit for the unpolarized case,
and the results may increase by a factor of $\rm 1+P_e$ if the polarized electron beam is considered,
where $\rm P_e$ is the degree of the longitudinal polarization of the beam.
We take $\rm P_e = 0.8$ as the default value.
In our numerical calculation, the SM inputs are
$\rm \alpha_{M_Z}=1/127.9$, $\rm G_f=1.1663787\times 10^{-5}GeV^{-2}$, $\rm \alpha_s=0.1182$, $\rm M_Z=91.1876GeV$ and $\rm M_h=125.09GeV$.
For the GM model inputs we take
$\rm \sin\alpha =0.303$, $\rm M_H =288\ GeV$, $\rm M_{H_3} =304\ GeV$, $\rm M_1=100\ GeV$, $\rm M_2 =100\ GeV$,
while let $\rm \sin\theta_H$ run in the range (0.1, 0.8) and $\rm M_{H_5}$ in the range $\rm (200\ GeV,\ 800\ GeV)$.
Typical fixed value of $\rm \sin\theta_H=0.3$ and $\rm M_{H_5} =300\ GeV$ is chosen as the benchmark point if there is no other statement.
We adopt the basic kinematical cuts as follows:
\begin{eqnarray} \nonumber
&&\rm  P_T^{\ell}>10\ GeV,\ \ \rm P_T^{jet}>20\ GeV,\ \ \rm E_T^{miss} >10\ GeV,\\  \nonumber
&&\rm  \eta^{\ell}<2.5, \ \  \rm \eta^{j}<5,\\
&&\rm  \Delta R_{jj}>0.4 ,\ \ \rm \Delta R_{j\ell}>0.4, \ \ \rm \Delta R_{\ell\ell}>0.4,
\label{basiccuts}
\end{eqnarray}
where $\rm P_T^{\ell}$ and $\rm P_T^{jet}$ are transverse momentum of leptons and jets respectively.
$\rm E_T^{miss}$ is the missing transverse momentum. $\rm \Delta R = \sqrt{\Delta \Phi^2 + \Delta \eta^2}$ is the separation
in the rapidity($\rm \eta$)-azimuth($\rm \Phi$) plane. The cuts are defined in the lab frame.

\begin{figure}[hbtp]
\centering
\includegraphics[scale=0.55]{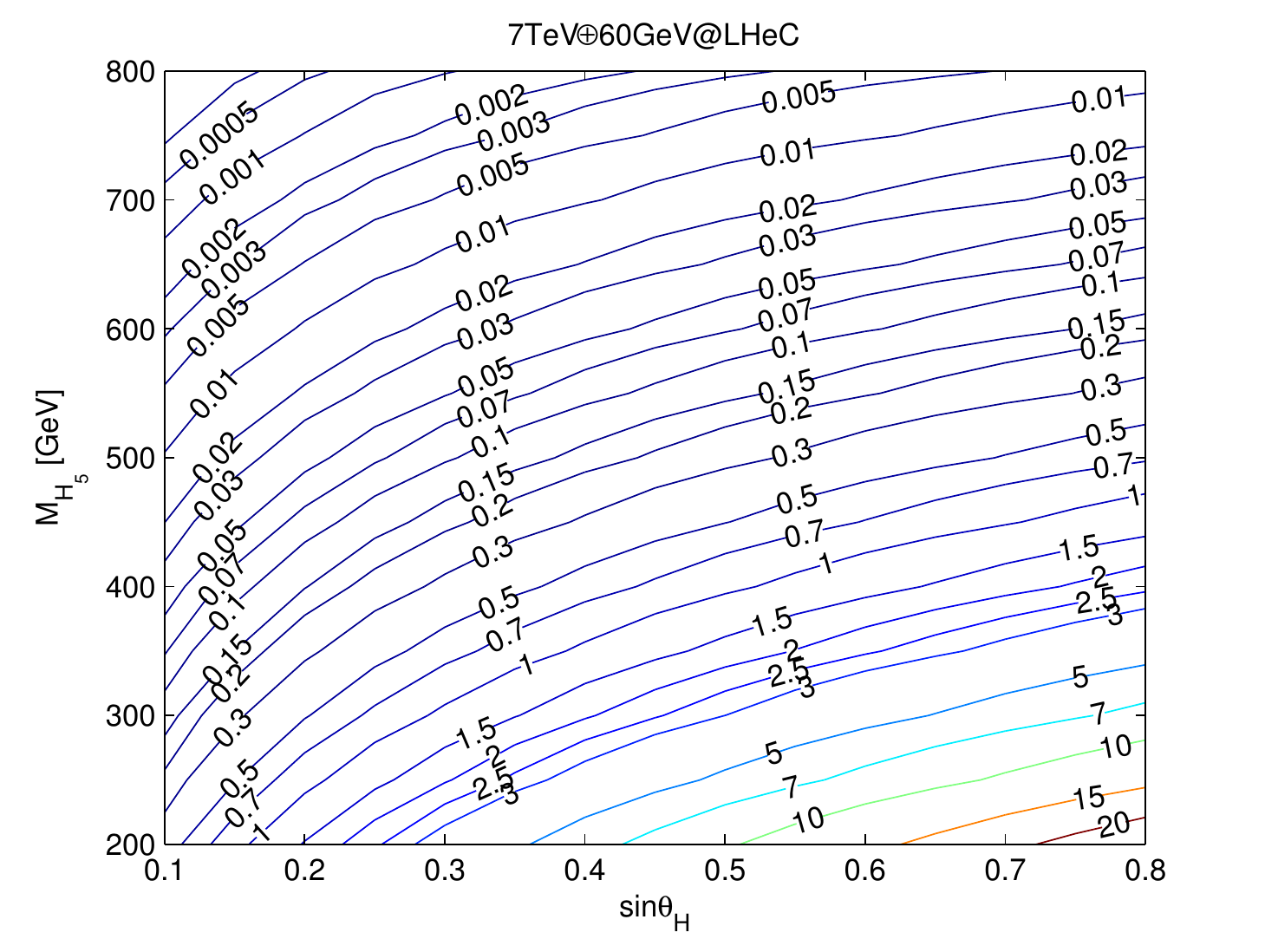}
\includegraphics[scale=0.55]{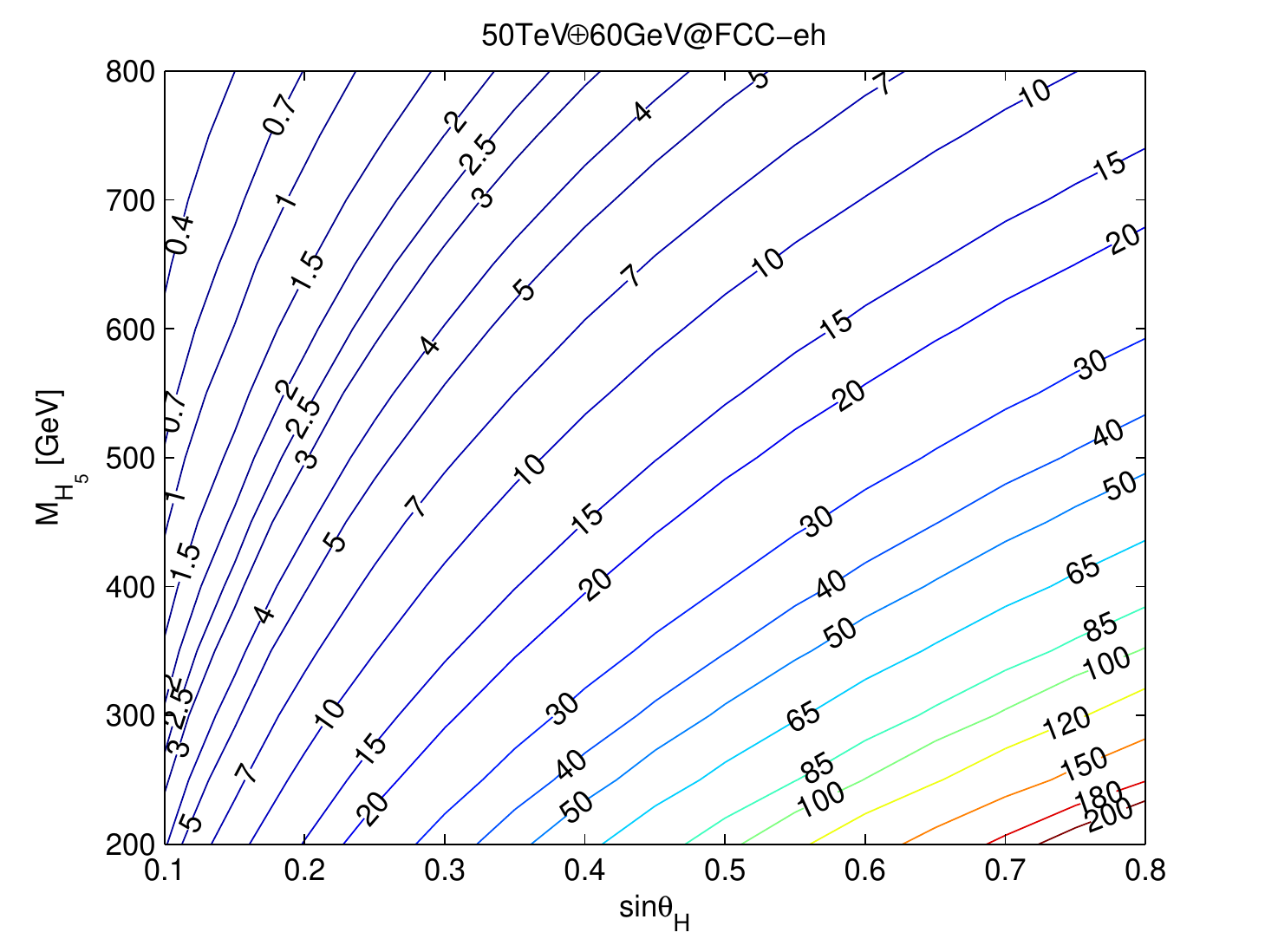}
\caption{\label{GM parameter scan}
Total cross sections (in $\rm fb$) for doubly-charged 5-plet Higgs production
$\rm e^-p\to\nu_e H_5^{--}j$ as the functions of $\rm sin\theta_H$ and $\rm M_{H_5}$.
Cut of $\rm P_T^{jet}>20\ GeV$ is taken. Left panel and right panel correspond to the 7 TeV LHeC and 50 TeV FCC-eh, respectively.
The electron beam is 60 GeV.}
\end{figure}
Before doing the full signal and background simulation,
we present a general VBF production cross section of the doubly-charged Higgs boson $\rm e^-p\to\nu_e H_5^{--}j$
at the LHeC and FCC-eh, in order to have a basic idea of its
production rate, with the proton beam energy chosen to be 7(50) TeV
and the electron beam is 60 GeV as proposed.
The total cross sections are plotted in Fig.\ref{GM parameter scan}
in units of fb as functions of $\rm \sin\theta_H$ and $\rm M_{H_5}$.
As shown in the figures, the cross sections enhance obviously when $\rm \sin\theta_H$ is becoming larger,
while on the other hand, suppressed as $\rm M_{H_5}$ become larger, due to the suppression of the phase space.
Furthermore, the production rate of 50 TeV FCC-eh is approximate 10 times larger than that of 7 TeV LHeC
in the large $\rm \sin\theta_H$ low (and middle) $\rm M_{H_5}$ range,
while 100 times larger in the low $\rm \sin\theta_H$ large $\rm M_{H_5}$ range.
Notice the allowed boundary (blue curve) in Fig.\ref{allowed parameter space} should be taken into account,
thus the cross section should be constrained correspondingly.

\subsection{Selections and Discovery Potential}

\begin{figure}[hbtp]
\centering
\includegraphics[scale=0.4]{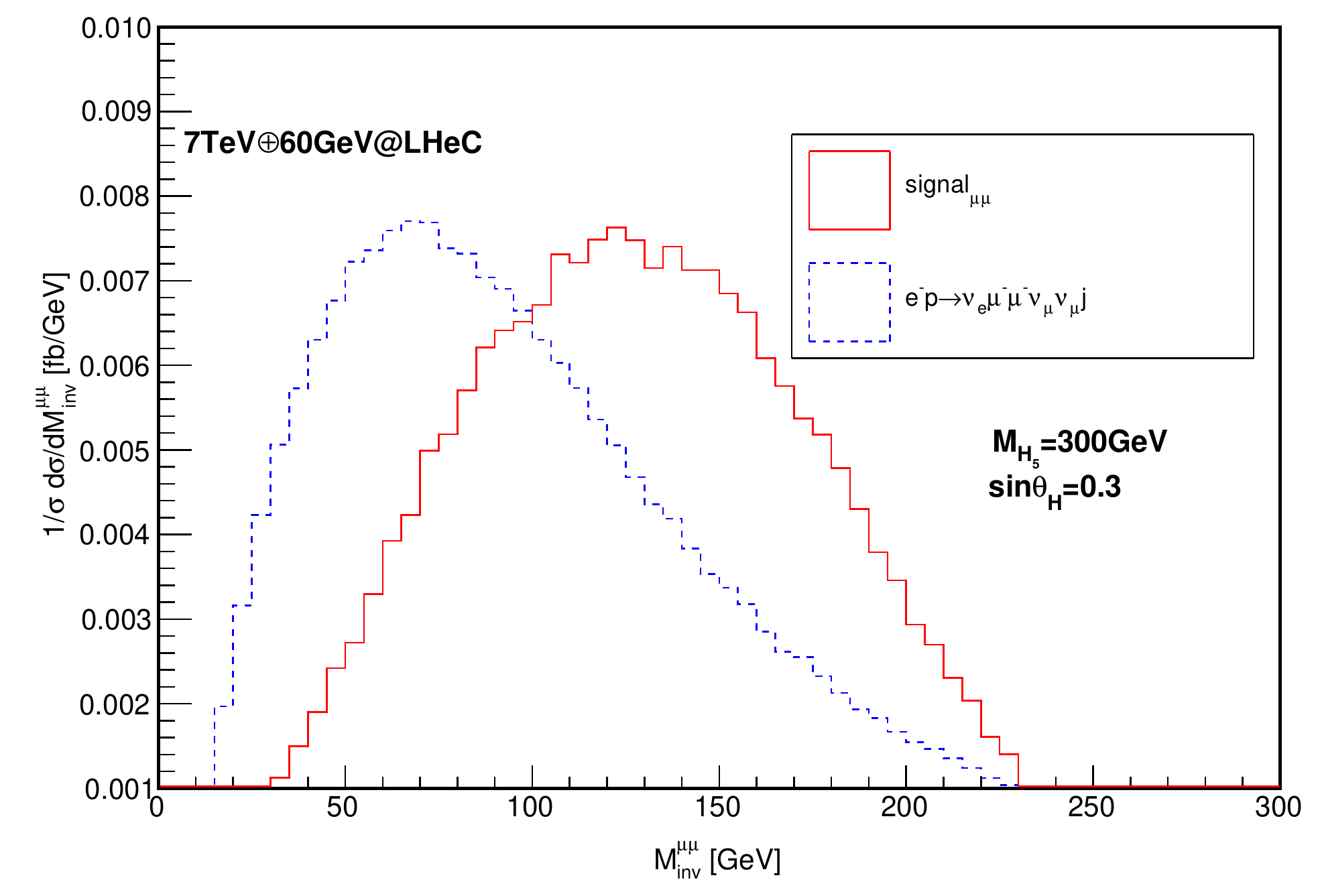}
\includegraphics[scale=0.4]{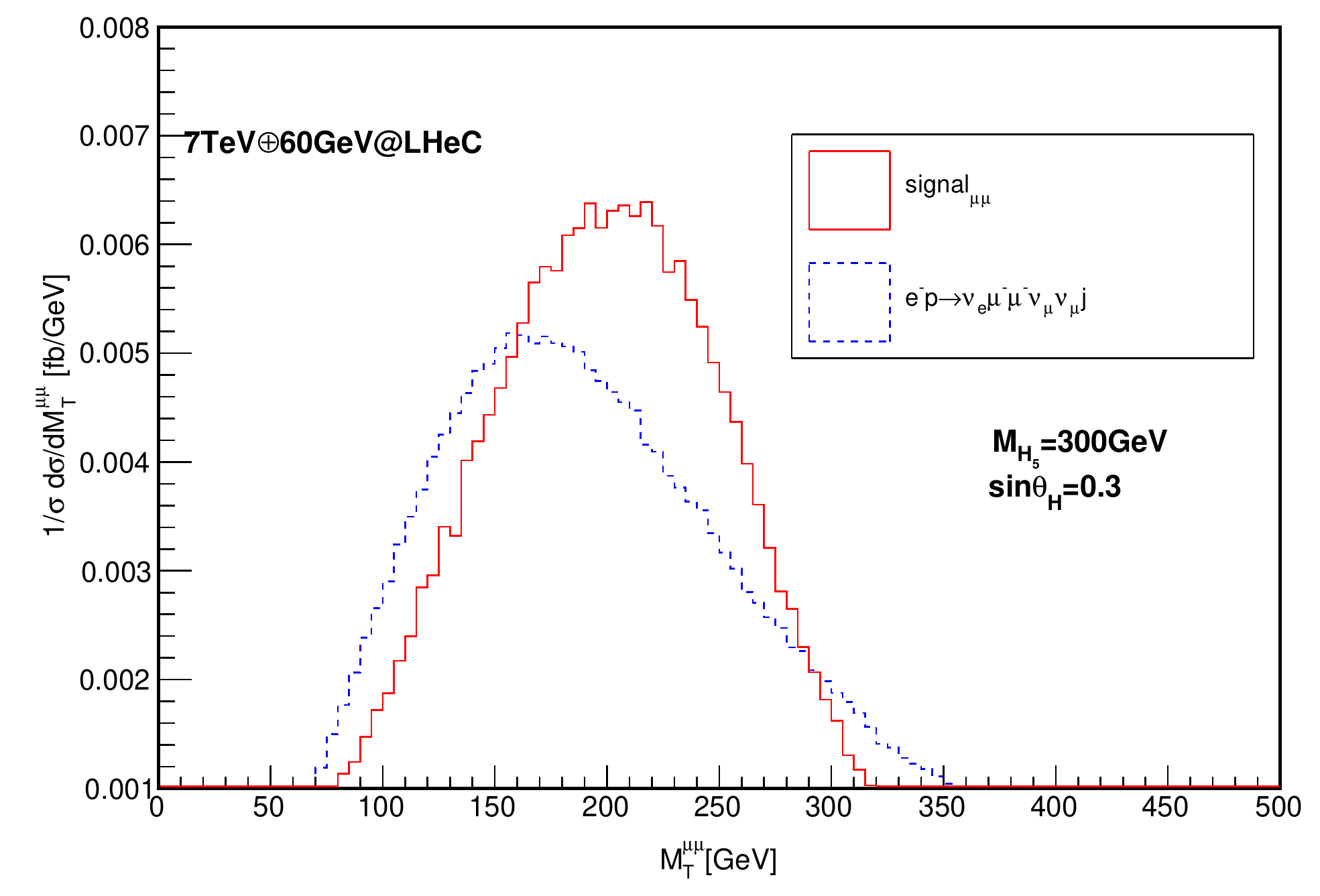}\\
\includegraphics[scale=0.4]{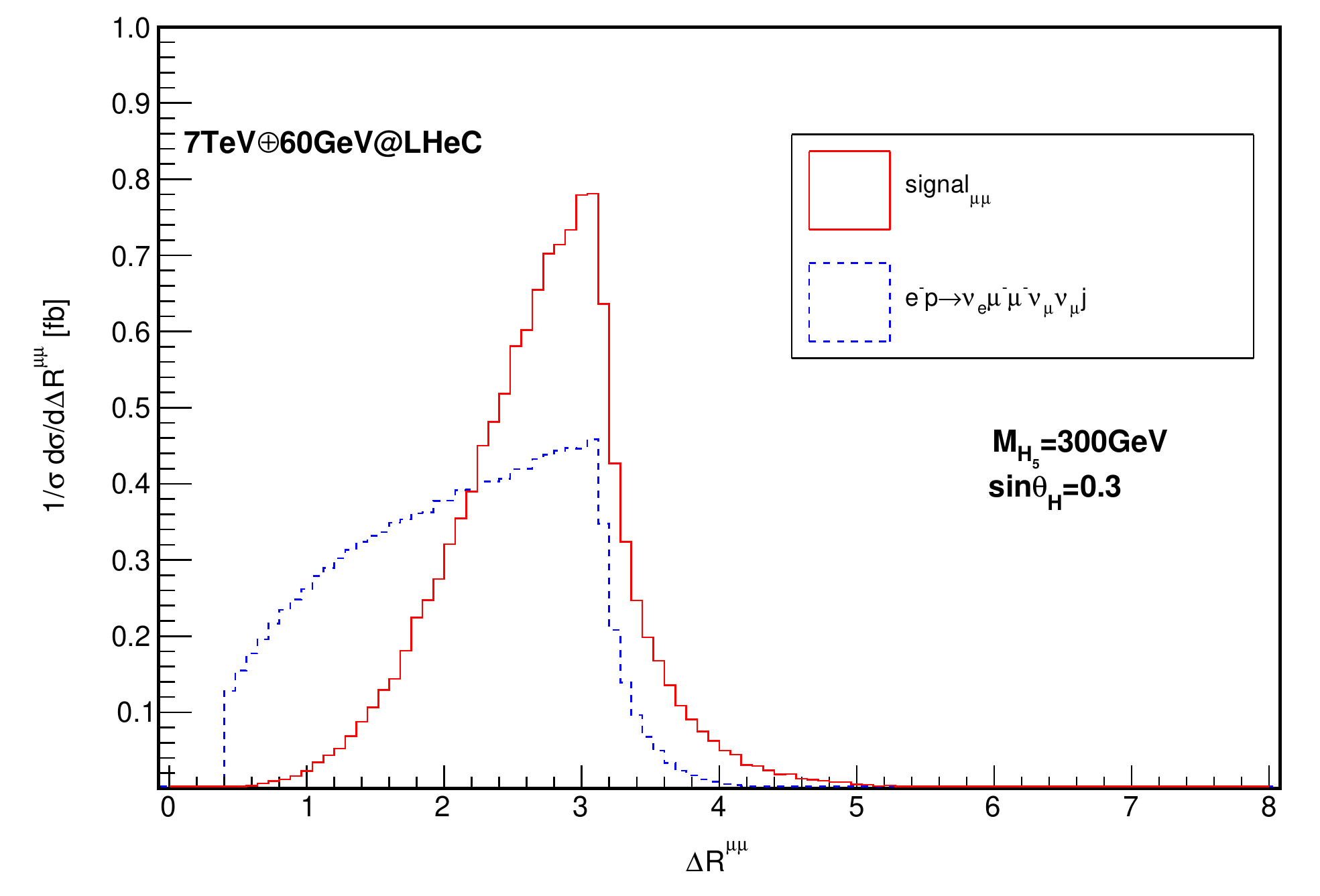}
\includegraphics[scale=0.4]{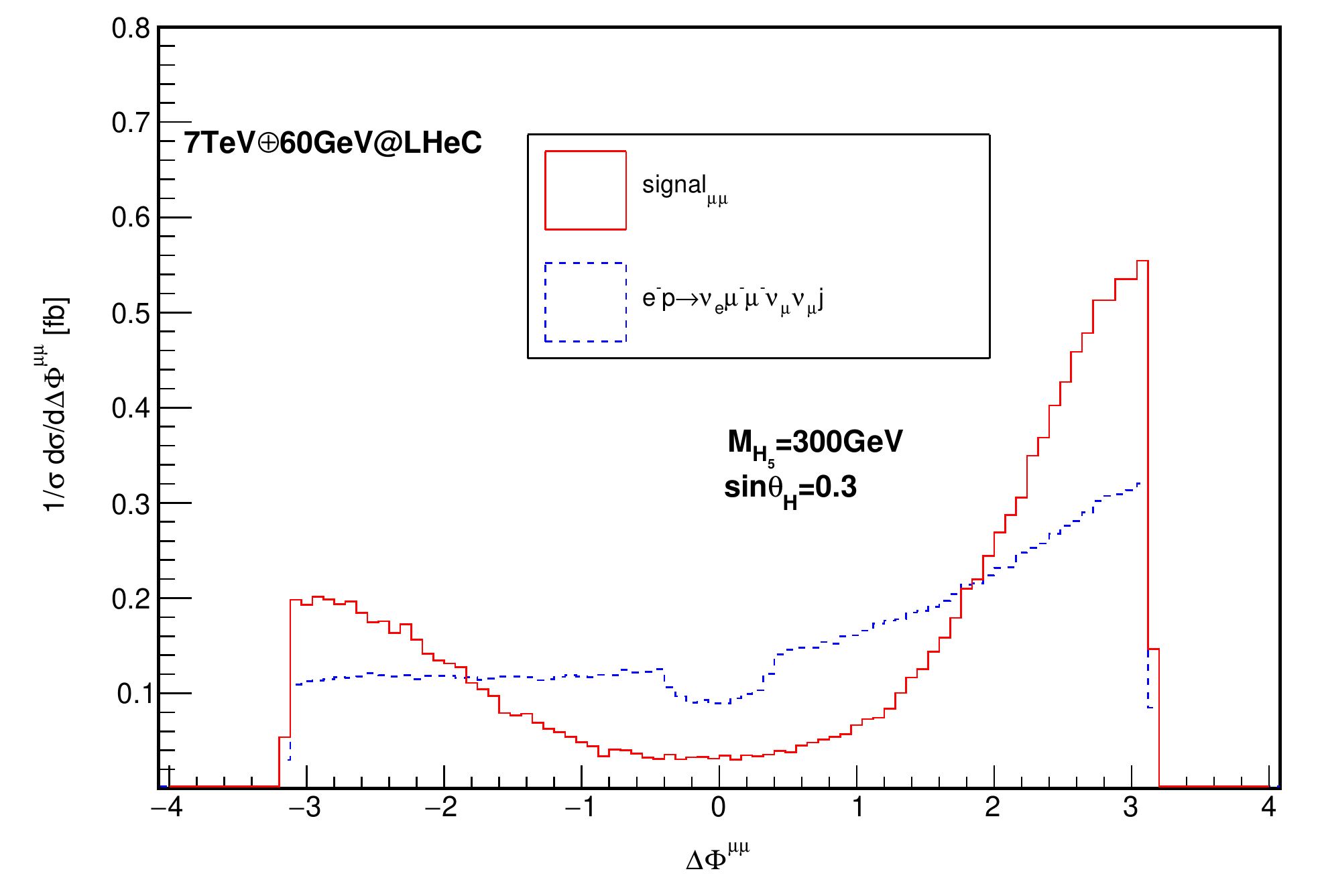}
\caption{\label{7_distributions}
Various kinematical distributions (in units of fb per bin)
for the signal and backgrounds at the 7 TeV LHeC. The electron beam is 60 GeV.
Here $\rm M_{H_5}=300\ GeV$ and $\rm \sin\theta_H=0.3$. Plots are unit normalized.}
\end{figure}

Now let's study the signal and the background at the distribution level.
After adopting the basic cuts in Eq.(\ref{basiccuts}), our sample selection is simply
\begin{eqnarray}
\rm 2\ \mu^- + E^{miss}_T + \geq 1\ jet(s).
\end{eqnarray}
Taking the typical benchmark input for the signal ($\rm \sin\theta_H=0.3$ and $\rm M_{H_5} =300\ GeV$),
the expected number of events is about 10.5(10.2) for 1$\rm ab^{-1}$ LHeC (100$\rm fb^{-1}$ FCC-eh),
as presented in Table.\ref{SBaftercuts}.
As we said, the only background is simply $\rm e^-p\to\nu_e \mu^-\mu^-\nu_\mu\nu_\mu j$,
including contributions dominant by the 2-$\rm W^-$ resonant ones.
The corresponding background events are about 10.5(8.3).
In fact, we are already very lucky since in this case the signal is
almost the same order compare to the background.
\begin{table}[htbp]
\begin{center}
\begin{tabular}{c | c  c  c  c  c  }
\hline
\makecell {7TeV$\oplus$60GeV$@$LHeC \\ $[\rm 1\ ab^{-1}]$ } & \makecell{ same-sign \\ $\mu^-\mu^-$}
& \makecell { $\rm \Delta R^{\mu\mu}$ \\ > 1.92 }  & \makecell{ $\rm \Delta\Phi^{\mu\mu}\in (-\pi, -1.28)$\\ or\ $(1.36,\pi)$ }
&  \makecell{ $\rm M_{inv}^{\mu\mu}$\\ >75\ GeV }   &   \makecell{  $\rm M_{T}^{\mu\mu}$ \\ >40\ GeV  } \\
\hline
$\rm signal_{\mu\mu}$[$\rm \sin\theta_{H}=0.3$]         &  10.46  & 9.45   & 9.24  & 9.21  &  10.23    \\
\hline
$\rm B$[$\rm e^-p\to\nu_e \mu^-\mu^-\nu_\mu\nu_\mu j$]  & 10.52    & 6.43  & 7.13  & 6.71  & 9.58  \\
\hline
$SS$                                            &  2.84    & 3.13  & 2.96   & 3.02  & 2.89  \\
\hline\hline
\makecell {50TeV$\oplus$60GeV$@$FCC-eh \\ $[\rm 100\ fb^{-1}]$ }  & \makecell{ same-sign \\ $\mu^-\mu^-$}
& \makecell { $\rm \Delta R^{\mu\mu}$ \\ > 1.6 }  & \makecell{ $\rm \Delta\Phi^{\mu\mu}\in (-\pi, -0.88)$\\ or\ $(1.04,\pi)$ }
&  \makecell{ $\rm M_{inv}^{\mu\mu}$\\ >70\ GeV }   &   \makecell{  $\rm M_{T}^{\mu\mu}$ \\ >78\ GeV  } \\
\hline
$\rm signal_{\mu\mu}$[$\rm \sin\theta_{H}=0.3$]&  10.16  & 9.44 & 9.25  &  9.13 &   9.75   \\
\hline
$\rm B$[$\rm e^-p\to\nu_e \mu^-\mu^-\nu_\mu\nu_\mu j$]  & 8.29  & 5.88  & 6.41  & 5.40  & 7.04  \\
\hline
$SS$                  &  3.03  & 3.23 &  3.08  & 3.24  & 3.11  \\
\hline
\end{tabular}
\caption{\label{SBaftercuts}
Expected number of events and signal significance ($SS$)
evaluated with $\rm 1\ ab^{-1}$ integrated luminosity at the LHeC and  $\rm 100\ fb^{-1}$ at the FCC-eh.
Here $\rm M_{H_5}=300\ GeV$ and each time take only one cut, so that here is not the cut flow.}
\end{center}
\end{table}
We will aim at finding, maybe, more efficient selections in order to allow the
best separation between noise-related and signal-related events,
and show better search strategies for the $\rm H_5 ^{--}$ at the ep colliders.
We plot various kinematical distributions for the signal and backgrounds,
including the missing transverse momentum distribution ($\rm E_{T}^{miss}$),
the scalar sum of transverse momenta distributions ($\rm H_T$),
the leading and next-to-leading $\mu$-lepton transverse momentum distribution ($\rm p_T ^{\mu}$),
the vector sum of the transverse momenta of all visible particles distribution ($\rm p_T ^{vis}$), etc.
However, what make us feel more interested, are those distributions from the same-sign $\mu^-$ system.
For example, the azimuthal angle separation $\Delta \Phi$,
the rapidity-azimuth plane separation $\rm \Delta R$, the invariant mass $\rm M_{inv}$,
the $\rm p_T$ distributions for the same-sign $\mu$ system,
as well as the the transverse mass $\rm M_T^{\mu\mu}$ distribution, which is defined by
\begin{eqnarray}
\rm M_T^2 ={\bigg[\sqrt{M_{inv}^2 + {(p_T^{vis})}^2 } +|/\kern-0.57em p_T | \bigg]}^2 -{\Big[ p_T^{vis} + /\kern-0.57em p_T \Big]}^2,
\end{eqnarray}
where $\rm /\kern-0.57em p_T$ is the missing transverse momentum determined
by the negative sum of visible momenta in the transverse direction.
We present some of these distributions of the same-sign $\mu$ system in Fig.\ref{7_distributions}.
The red solid curve is for the signal and blue dashed one is for the background.
The plots are unit normalized for both the signal and background
and the results are for the 7 TeV LHeC, while for the 50 TeV FCC-eh we can get similar ones thus not shown.
From the distributions of the same-sign $\mu$ system in Fig.\ref{7_distributions},
we do find some obvious separation between the signal and the background, as expected.
For example, for $\rm M_{inv}^{\mu\mu}$ distribution, the peak of the signal appears at the position with the value
larger than that of the background, thus can be easily separated.
There is also one interesting thing that if we increase $\rm M_{H_{5}}$,
the separation between these two peaks will increase, thus leading better signal vs background separation.
This feature doesn't make sense for $\rm \Delta R^{\mu\mu}$ and $\Delta \Phi^{\mu\mu}$ distributions,
though they have already shown some other nice features.
This is indeed the reason why when we are facing larger value of $\rm M_{H_{5}}$, the optimized
selection is $\rm M_{inv}^{\mu\mu}$ instead of the others.
When $\rm M_{H_{5}}$ is not very large, for example, 300 GeV in Table.\ref{SBaftercuts},
the optimized selection can be $\rm \Delta R^{\mu\mu}$.
In this case, the signal peaked strongly around $\rm \Delta R^{\mu\mu}=3$,
while the background spreads in the range one to four, and being much flatter.
The behavior of the $\Delta \Phi^{\mu\mu}$ distribution is also very interesting.
We can see that the signal is suppressed in the middle range of the absolute value of $\Phi_{\mu\mu}$ ($|\Delta \Phi_{\mu\mu} |$),
while enhanced obviously with the increase of $|\Delta \Phi_{\mu\mu}|$.
Furthermore, the peak for $\Delta \Phi_{\mu\mu} > 0$ is much larger than that for $\Delta \Phi_{\mu\mu} < 0$.
On the other hand, the corresponding background spread evenly in the whole range $\Delta \Phi_{\mu\mu} \in (-\pi, \pi)$
with a much flat curve.
The reason of the different behavior between the signal and the background distributions are indeed straight forward,
as we may have already pointed out at some points.
The same-sign $\mu$ system comes from same-sign W boson pair.
For the signal, $\rm W^-W^-$ is derived from the doubly charged Higgs $\rm H_5 ^{--}$ through s-channel resonance,
which is different from the background that the same-sign W pair is derived from the dominant contributions
origin from four-vector boson vertexes.
Based on the difference between the signal and the background distributions, we make a comparison
in Table.\ref{SBaftercuts}, in order to find out the optimized selection.
The expected number of events and the corresponding signal significance (SS) are
evaluated with 1$\rm ab^{-1}$ integrated luminosity at the LHeC and 100$\rm fb^{-1}$ at the FCC-eh.
The significance is calculated by the following formula:
\begin{eqnarray}
\rm {\cal{SS}} = \sqrt{2[(n+b)\log(1+\frac{n}{b})-n]}
\end{eqnarray}
where n is the number of events and b is the number of backgrounds evaluated with the corresponding integrated luminosity.
Notice here $\rm M_{H_{5}}$ equal 300 GeV, not very large, that's why the optimized selection is $\rm \Delta R^{\mu\mu}$,
but not $\rm M_{inv}^{\mu\mu}$ instead.
In Table.\ref{SBaftercuts} and also our calculation,
each time we take only one cut (not the cut flow), so as not to cut the events too much,
otherwise the significance can be reduced.
What is the best cut and corresponding significance,
are determined in a somehow automatic way, relying on the machine computation.

\begin{figure}[hbtp]
\centering
\includegraphics[scale=0.4]{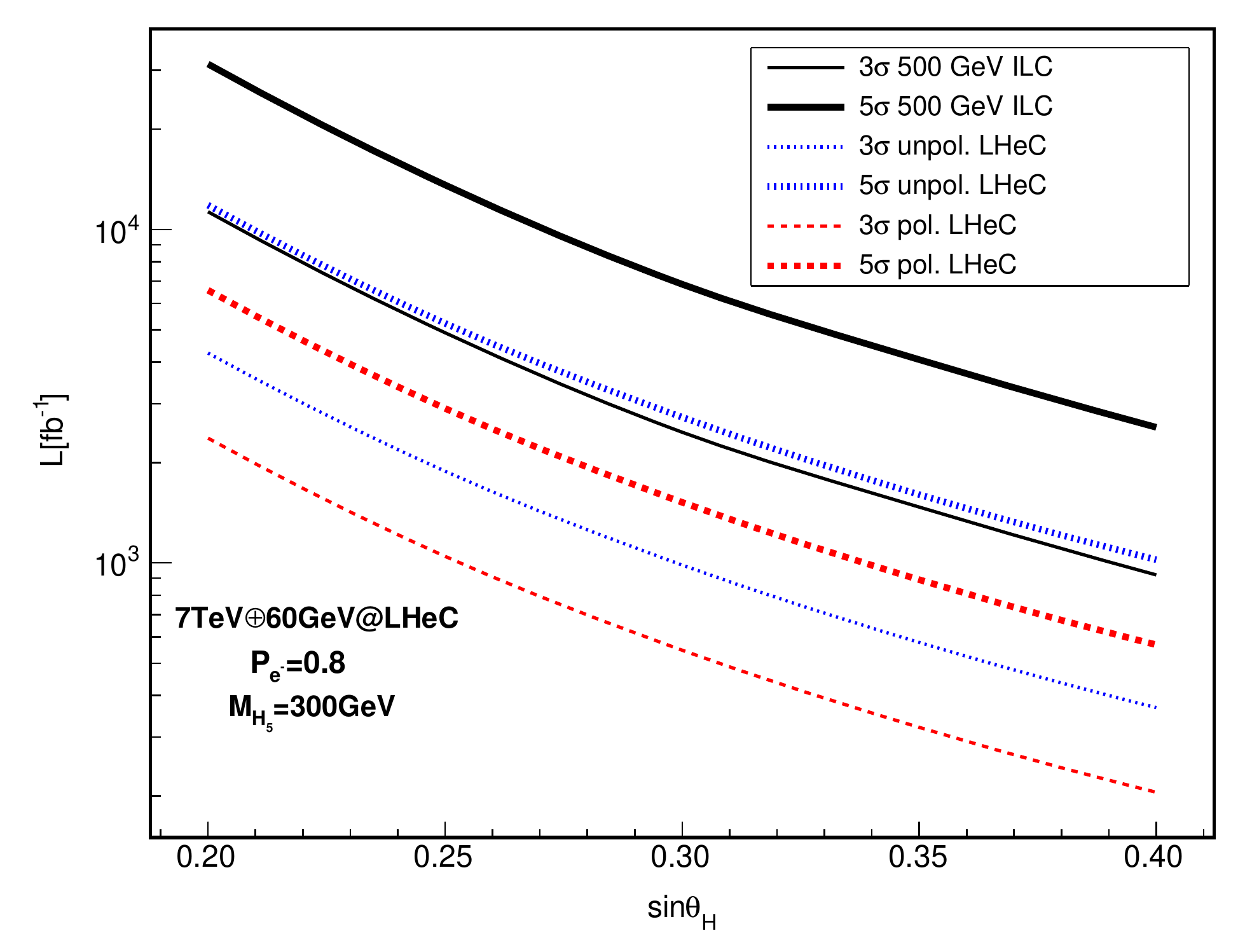}
\includegraphics[scale=0.4]{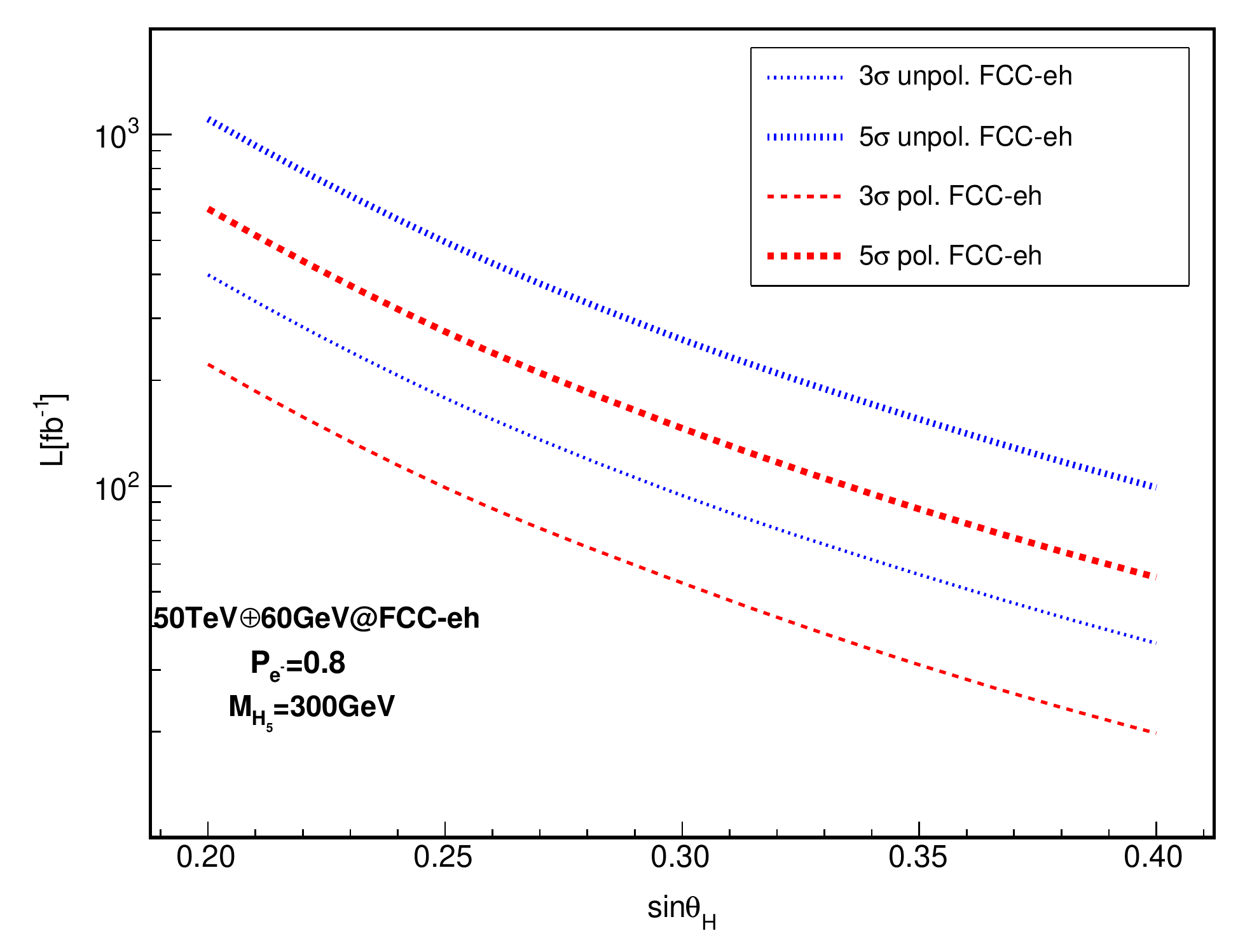}
\vspace{-0.3cm}
\caption{\label{luminosity}
The lowest necessary luminosity with $3\sigma$ and $5\sigma$
discovery significance as a function of $\rm sin\theta_H$.
Here we use $10\%$ systematic uncertainty for background yields only.}
\end{figure}
In Fig.\ref{luminosity} we calculate the lowest necessary luminosity (L)
with $3\sigma$ and $5\sigma$ discovery significance as a function of
$\rm sin\theta_H$. $\rm M_{H_{5}}$ is fixed to be 300 GeV.
We present the results for the 7 TeV LHeC (left panel) and 50 TeV FCC-eh (right panel)
with both the unpolarized and polarized ($\rm p_{e^-}=0.8$) electron beams.
The dashed and dotted curves correspond to polarized and unpolarized, respectively.
We use $10\%$ systematic uncertainty for background yields only.
It is shown that high integrated luminosity is required to probe small $\rm sin\theta_H$.
At the FCC-eh, the lowest necessary luminosities are much reduced than that at the LHeC as expected.
In addition, the polarized beam could make the measurement even better.
The results are compared with a related study
at the International Linear Collider (ILC) in Ref.\cite{ILC_sinh},
see the black solid curve in Fig.\ref{luminosity}.
Here we add a comment that the ILC results in the $\rm \sin\theta_H-L$ plane
are obtained from the charged fiveplet scalars, but not the doubly-charged sector.
A much closed comparison would be the doubly-charged scalar production
at the like-sign lepton colliders through similar production mechanisms,
which is currently being working on. Nevertheless the parameters are the same,
and we find that the ep colliders show some better features to detect them.
\begin{table}[htbp]
\begin{center}
\begin{ruledtabular}
\begin{tabular}{c | c  c  c  c  c  c c c c  c}
\multirow{2}{*}{$\cal{L}$[$\rm fb^{-1}$]}  & \multicolumn{2}{c}{ILC\cite{ILC_sinh}}    & \multicolumn{2}{c}{unpol. LHeC} & \multicolumn{2}{c}{pol. LHeC}
& \multicolumn{2}{c}{unpol. FCC-eh} & \multicolumn{2}{c}{pol. FCC-eh} \\
  & $3\sigma$   & $5\sigma$    & $3\sigma$   & $5\sigma$    & $3\sigma$   & $5\sigma$ & $3\sigma$   & $5\sigma$ & $3\sigma$   & $5\sigma$ \\
\hline
$\rm sin\theta_H=0.20$  &  11300  &  31350  &  4259  & 11830   &  2366 & 6572  & 399  & 1107 &  222 & 615  \\
$\rm sin\theta_H=0.25$  &  4900   &   13600 &  1883  &  5229   &  1046 &  2905 & 178  &495   &  99  & 275    \\
$\rm sin\theta_H=0.30$  &  2465   &   6850  &  985   & 2734    &  547  & 1519  & 94   &261   &  53  & 146 \\
$\rm sin\theta_H=0.35$  &  1470   &   4070  &  577   & 1602    &  321  &  890  & 56   &155   &  31  & 86   \\
$\rm sin\theta_H=0.40$  &  920    &   2550  &  368   &  1022   & 205   &  568  & 35.7 &99    &  19.8  &55    \\
\end{tabular}
\end{ruledtabular}
\begin{minipage}{16cm}
\caption{\label{lumi_compare}
The lowest necessary luminosity with $3\sigma$ and $5\sigma$ discovery significance correspond to the plot in Fig.\ref{luminosity}.}
\end{minipage}
\end{center}
\end{table}
In order to be clear, the corresponding values of the luminosity are presented in Table.\ref{lumi_compare},
we can thus confirm the same conclusion straightly.
\begin{table}[htbp]
\begin{center}
\begin{tabular}{c | c c c c c}
\hline\hline
\multirow{2}{*}{$\cal{L}$[$\rm fb^{-1}$]}  & \multicolumn{2}{c}{unpol. FCC-eh} & & \multicolumn{2}{c}{pol. FCC-eh} \\
  &  $3\sigma$   & $5\sigma$ & & $3\sigma$   & $5\sigma$ \\
\hline
$\rm M_{H_5}=400$  &  112.6   & 312.6  &  & 62.6  & 173.7  \\
$\rm M_{H_5}=500$  &  2355   & 6540  &  & 1308  &  3634   \\
$\rm M_{H_5}=600$  &  7925   &  22014 &  & 4403  & 12230 \\
\hline\hline
\end{tabular}
\begin{minipage}{16cm}
\caption{\label{lumi_compare_2}
The lowest necessary luminosity with $3\sigma$ and $5\sigma$ discovery significance for different values of $\rm M_{H_5}$ where $\rm sin\theta_H=0.4$.}
\end{minipage}
\end{center}
\end{table}
In Table.\ref{lumi_compare_2} we fix $\rm sin\theta_H$ equal 0.4 and shift the value of $\rm M_{H_5}$,
the lowest necessary luminosity is presented. As the mass increase,
the signal production rate is strongly suppressed
due to the exchange of the s-channel heavy doubly-charged Higgs,
thus making the detection much more difficult.

\begin{figure}[hbtp]
\includegraphics[scale=0.6]{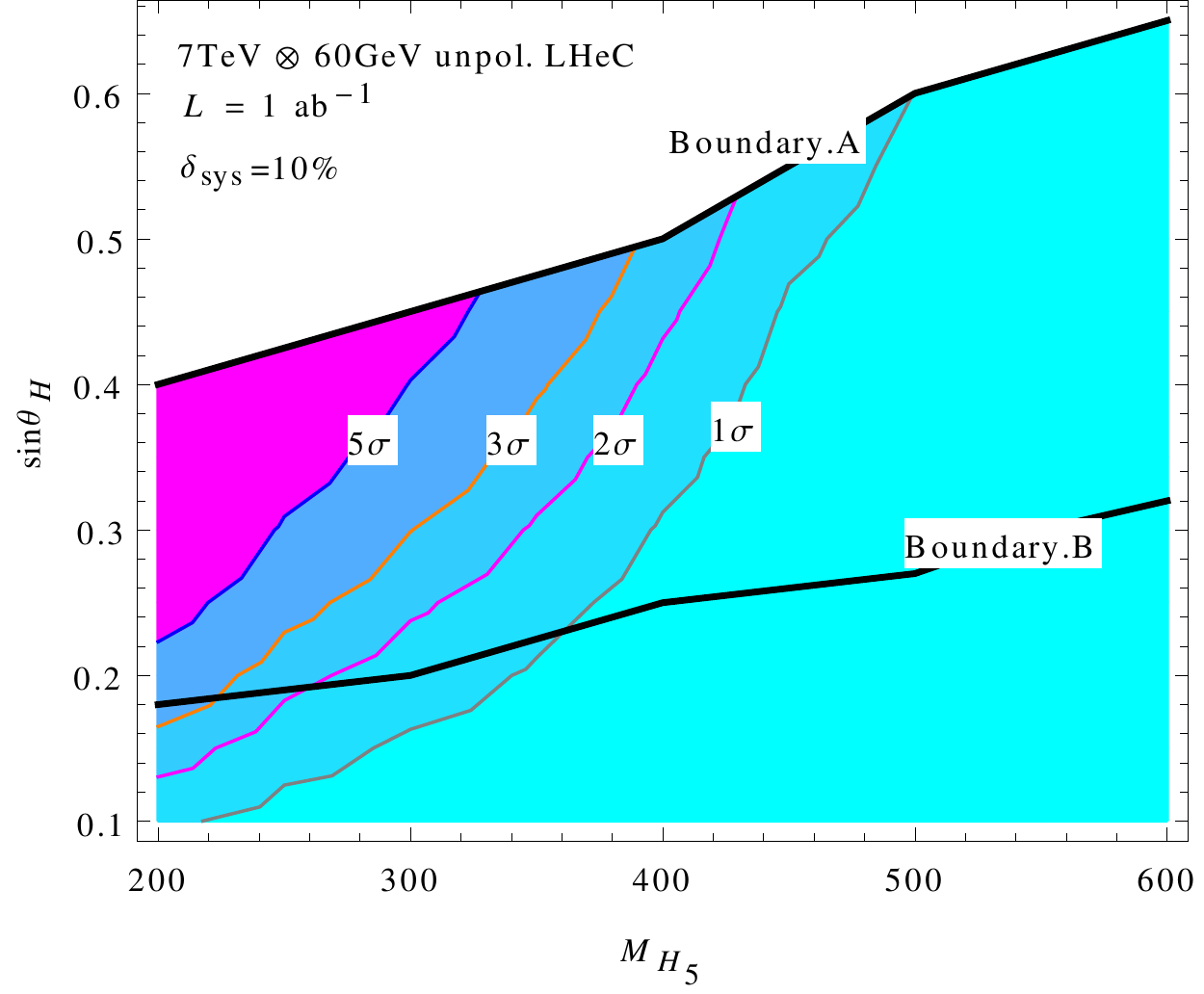}
\includegraphics[scale=0.6]{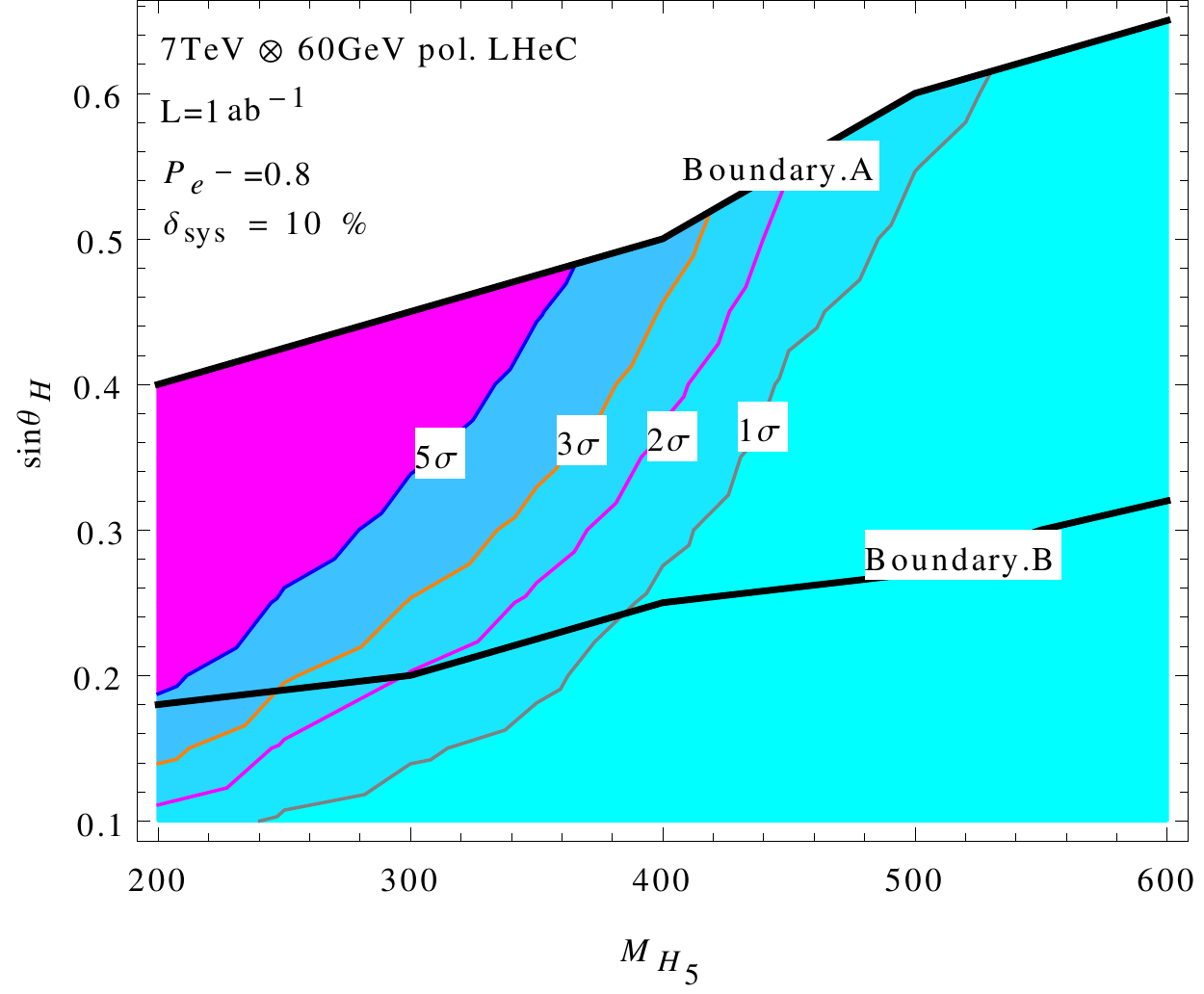}\\
\includegraphics[scale=0.6]{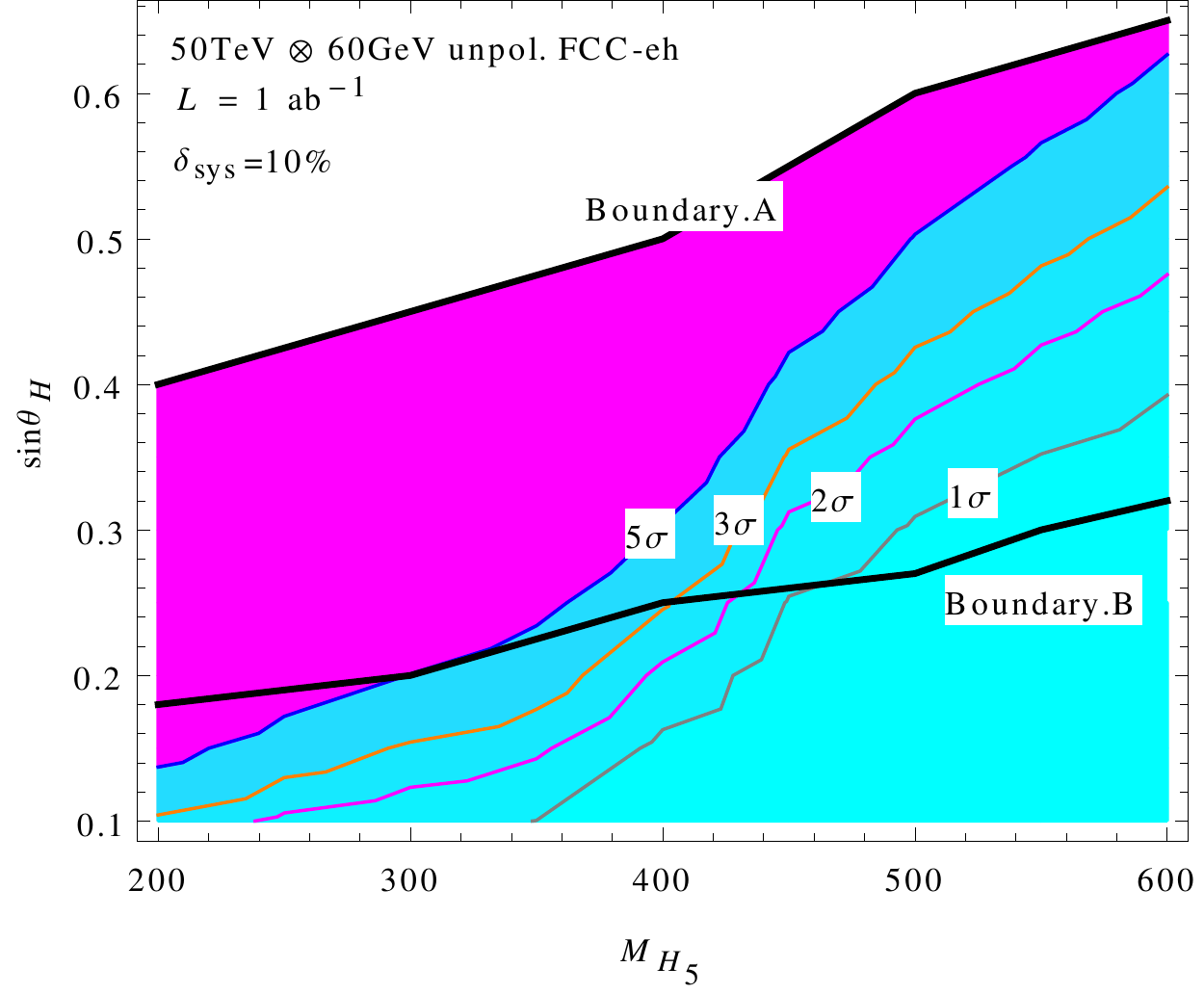}
\includegraphics[scale=0.6]{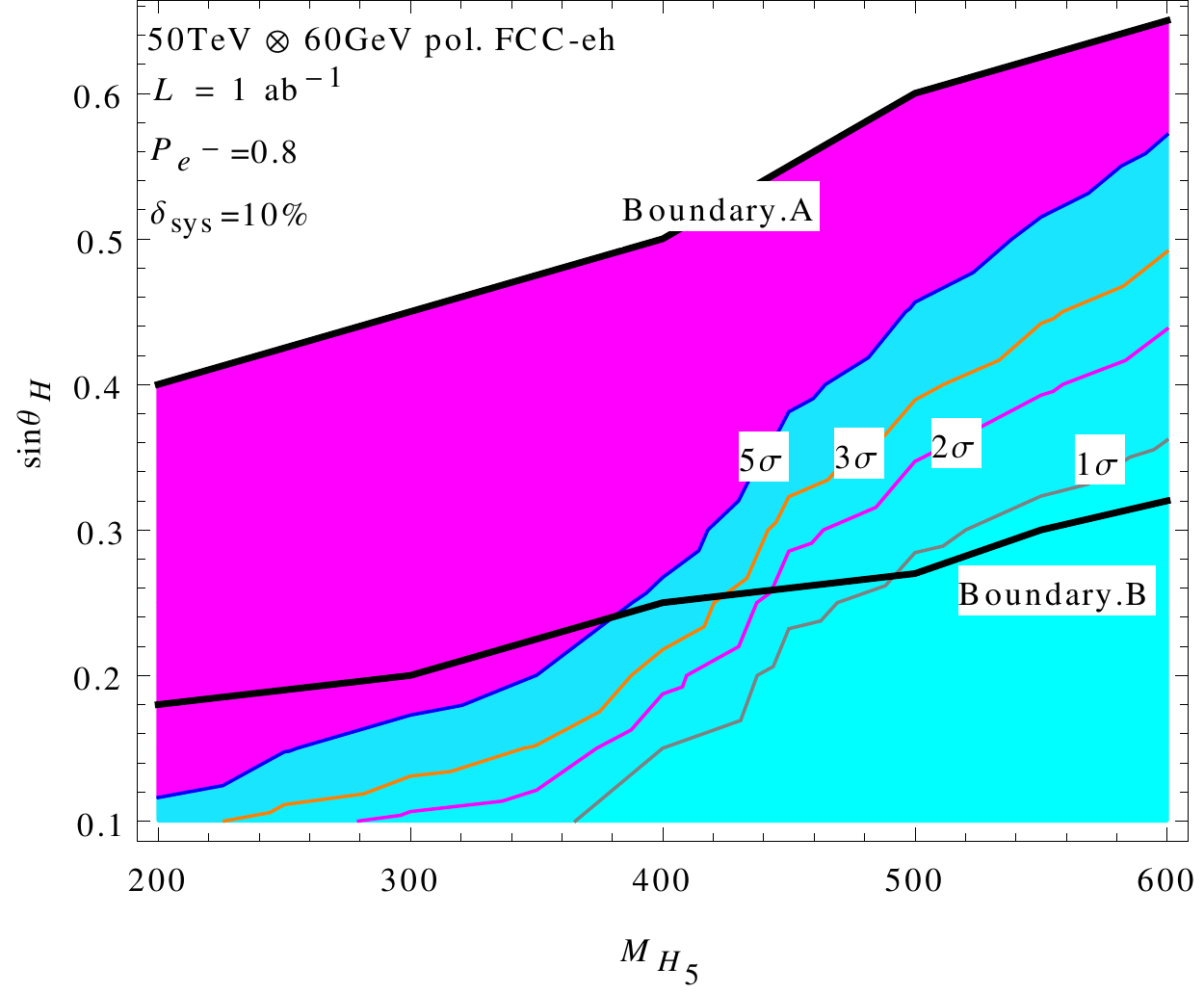}
\caption{\label{epH5potential}
The discovery contour with respect to $\rm sin\theta_H$ and $\rm M_{H_5}$ by considering
signal significance based on events at the 7TeV LHeC and 50 TeV FCC-eh.
The proposed luminosity is fixed to be 1$\rm ab^{-1}$. $\rm P_e = 0.8$ for the electron beam.
$10\%$ systematic uncertainty for background yields only is used.}
\end{figure}
Finally in Fig.\ref{epH5potential} we present the discovery contour
with respect to $\rm sin\theta_H$ and $\rm M_{H_5}$, by considering
signal significance based on events at the 7TeV LHeC and the 50 TeV FCC-eh.
The proposed luminosity is fixed to be 1$\rm ab^{-1}$.
We take $\rm P_e = 0.8$ for the electron beam.
Still, $10\%$ systematic uncertainty for background yields only is considered.
The gray, magenta, orange and blue solid lines correspond to
1$\sigma$, 2$\sigma$, 3$\sigma$, and 5$\sigma$ regions, respectively.
Let's see the first panel in Fig.\ref{epH5potential}
which is the contour plot for the unpolarized 7 TeV LHeC.
For a fixed value of $\rm \sin\theta_H$, as the masses increase, the signal significance is reduced.
Similarly, when we fix $\rm M_{H_5}$, the signal significance decreases as $\rm \sin\theta_H$ becoming small.
The upper black solid curve is the boundary (Boundary.A)
corresponding to the blue line in Fig.\ref{allowed parameter space}.
The lower one is the latest limit from the CMS experiment at the 13 TeV LHC
through electroweak production of same-sign W boson pairs in the two jet and two same-sign lepton final
state\cite{GM_Boundary_b}.
The parameter space above the boundary is excluded by the LHC experimental data.
The 5 sigma region is constrained in the space where $\rm M_{H_5}$ less than 300 GeV
and $\rm \sin\theta_H$ larger than 0.2. Unfortunately, this range is excluded by the latest CMS experiment.
When turn to the polarized LHeC, the discovery area can be a little enlarged, but the 5 sigma region is still excluded.
We see that at the LHeC, only a piece of small regions, where $\rm M_{H_5}$ less than 300 GeV
and $\rm \sin\theta_H$ less than 0.2, is allowed at the 2 and 3 sigma level.
The situation becomes better at the 50 TeV FCC-eh as can be see in the last two panels in Fig.\ref{epH5potential}.
For both the unpolarized and polarized case, some piece of 5 sigma region is still allowed.
The 2 sigma region can be enlarged up to more than 400 GeV. For $\rm M_{H_5}$ less than 400 GeV, FCC-eh collider
may show some possibilities to detect small $\rm \sin\theta_H$ regions.
There is one comment we want to add here. For our current study, we just perform the cut-and-count analysis
with only a few observables. The multi-variant analysis with more input observables may
greatly enhance the limits. More, for our study, we consider only the decay modes of
$\rm W^-W^-\to \mu^-\nu\mu^-\nu$. Other final state, like $\rm W^-W^- \to jj\mu^-\nu$
may also worth considering. Although these are beyond the scope of this paper,
there are still lots of potential to improve the limits at the ep colliders. The work related to them is also ongoing.

\section{CONCLUSION}

The possible existence of heavy exotic particles in the beyond Standard Models are highly expected.
The Georgi-Machacek (GM) model is one of such scenarios with an extended scalar sector
which can group under the custodial $\rm SU(2)_C$ symmetry.
There are 5-plet, 3-plet and singlet Higgs bosons under the classification
of such symmetry in addition to the SM Higgs boson.
In the GM model, there are 5-plets doubly-charged states so that the distinct phenomenological features should appear.
In this paper, we study the prospects for detecting the doubly-charged Higgs boson
($\rm H_5^{\pm\pm}$) through the vector boson fusion production at the electron-proton (ep) colliders.
We concentrate on our analysis through $\mu$-lepton pair production via pair of same-sign W bosons decay.
The $\rm H_5^{\pm\pm}$ discovery potentials at the ep colliders are presented.
Considering the most stringent limit from the LHC experiments,
the allowed parameter space to the production of doubly-charged Higgs boson is indeed strongly constrained.
It is found that at the LHeC, only a piece of small regions, where $\rm M_{H_5}$ less than 300 GeV
and $\rm \sin\theta_H$ less than 0.2, is allowed at the 2 and 3 sigma level.
While for $\rm M_{H_5}$ less than 400 GeV, the FCC-eh collider
may show some possibilities to detect small $\rm \sin\theta_H$ regions.

\section*{Acknowledgments} \hspace{5mm}
The author H. Sun would like to thank the comments and encourages
from the LHeC (Top$\&$Higgs$\&$BSM) working Group. 
Thanks for useful discussions with KeChen Wang and Georges Azuelos. 
This work is supported by the National Natural Science Foundation of China
(Grant No. 11675033), by the Fundamental Research Funds for the Central Universities
(Grant No. DUT15LK22).

\end{document}